\documentclass[psfig]{aa}

\input psfig.sty

\newcommand{\gsim}{\raisebox{-3.8pt}{$\;\stackrel{\textstyle >}{\sim}\;$}}
\newcommand{\lsim}{\raisebox{-3.8pt}{$\;\stackrel{\textstyle <}{\sim}\;$}}

\newcommand{\Msol}{$M_{\odot}$}

\newcommand{\etal}{\mbox{{\rm et~al.\ }}}

\def\smallskip{\vskip 6pt}

\begin{document}

\thesaurus{       }

\title {On radial gas flows, the Galactic Bar \\
and chemical evolution in the Galactic Disc} 

\author{L.\ Portinari and C.\ Chiosi }

\institute{
Department of Astronomy, University of Padova, Vicolo dell'Osservatorio 5, 
35122 Padova, Italy (portinari, chiosi@pd.astro.it)}

\offprints{L.\ Portinari}
\date{Received 23 August 1999/ Accepted 14 January 2000}

\maketitle
\markboth{Portinari and Chiosi: On radial gas flows, the Galactic Bar
and chemical evolution in the Galactic Disc}{}


\begin{abstract}

We develop a numerical chemical model allowing for radial flows of gas,
with the aim to analyse the possible role of gas flows in the chemical 
evolution of the Galactic Disc.
The dynamical effects of the Galactic Bar on the radial gas profile of the
Disc are especially addressed.

\end{abstract}


\section{Introduction}

The observed chemical and spectro--photometric properties of galaxies are
one of the main sources of information for our understanding
of galaxy formation and evolution. The corresponding theoretical modelling 
involves star formation (SF) as a basic ingredient. Unfortunately, this process
is rather poorly known on the large scales relevant to galaxy evolution.
Portinari \& Chiosi (1999, hereinafter PC99) analysed the effects of
adopting different SF laws in a chemical model for the Galactic Disc, a
system which we can study in great detail. In this
paper we address another phenomenon which can bear interesting effects on the 
chemical evolution of galaxies:
radial gas flows. A few papers in literature (Section~\ref{previous}) 
demonstrate that radial gas flows influence chemical models for the Disc, 
especially in their predictions on the metallicity gradient.
It is therefore interesting to discuss the radial profile of the Disc
with models including also radial flows, in addition to various options for
the SF law. In particular, radial flows can help to overcome some
difficulties that ``static'' models find in reproducing, at the same time,
the metallicity gradient and the radial gas profile of the Disc
(PC99).

We develop a chemical model with radial gas flows as a multi--dimensional
extension of the model of Portinari \etal (1998, hereinafter PCB98).
The model is described in Section~\ref{discrete} and in the appendices.
In Section~\ref{radfloweffects} we discuss the general qualitative effects 
of superposing radial flows upon a chemical model. In Section~\ref{bestfit} 
we present models 
for the Galactic Disc with radial gas flows and different SF laws, showing 
that radial flows provide an alternative or additional dynamical effect 
to the ``inside--out'' formation scenario to explain the metallicity gradient.
Section~\ref{bar} is dedicated to qualitative simulations of the dynamical 
effects of the Galactic Bar upon the gas distribution, with the aim to 
reproduce the molecular ring around 4 kpc, which static models cannot account 
for (PC99). Section~\ref{conclusions} contains a final summary and
conclusions.


\section{Radial flows: previous literature}
\label{previous}

The possibility that radial flows play a role in establishing the radial
metallicity gradients in galactic discs was first suggested by Tinsley \& 
Larson (1978). Following Lacey \& Fall (1985), we mention that radial gas 
flows in a disc can be driven by three main mechanisms:
\begin{enumerate}
\item
the infalling gas has a lower angular momentum than the circular motions in
the disc, and mixing with the gas in the disc induces a net radial inflow with
a velocity up to a few km~sec$^{-1}$;
\item
viscosity in the gas layer induces radial inflows in the inner parts of the 
disc and outflows in the outer parts, with velocities of 
$\sim$0.1~km~sec$^{-1}$;
\item
gravitational interactions between gas and spiral density waves
lead to large-scale shocks, dissipation 
and therefore radial inflows of gas (or outflows in the outer parts) with
typical velocities of $\sim$0.3~km~sec$^{-1}$ (e.g.\ Bertin \& Lin 1996 and
references therein); much larger velocities can be achieved in the
inner few kpc in the presence of a barred potential.
\end{enumerate}
In summary, radial flows are plausible with velocities of
$\sim$0.1--1~km~sec$^{-1}$, and they are expected to be inflows over most of 
the disc. Observational upper limits permit radial inflows in the Galactic 
Disc with velocities up to 5~km~sec$^{-1}$ at the present time.
For further details, see Lacey \& Fall (1985) and references therein.

The first of the above mentioned mechanisms was modelled in detail
by Mayor \& Vigroux (1980), and later by Pitts \& Tayler (1989, 1996),
Chamcham \& Tayler (1994). The effects of a generic inflow velocity profile 
on chemical evolution models 
has been explored by Lacey \& Fall (1985), Tosi(1988), G\"otz \& K\"oppen 
(1992), K\"oppen (1994), Edmunds \& Greenhow (1995). 
A different approach is that of viscous disc models which, rather than imposing
arbitrary radial velocity patterns, describe the
evolution of the gas distribution in the disc self--consistently, following
the model suggested by Lin \& Pringle (1987). Viscous chemical models
have been developed by Clarke (1989), Yoshii \& Sommer-Larsen (1989) and
Sommer-Larsen \& Yoshii (1990), Thon \& Meusinger (1998).
All these studies show how radial inflows can steepen the metallicity
gradients with respect to static models, especially if an outer cut--off of
SF is assumed.


\section{Modelling radial flows}
\label{discrete}

We formulate our chemical model with radial flows as a multi-dimensional
extension of the static model of PCB98 and PC99, an open model where the disc
forms gradually by accretion of protogalactic gas. The disc is divided 
in $N$ concentric rings or shells; in each ring $k$ the gaseous component 
and its chemical abundances evolve due to:
\begin{enumerate}
\item
depletion by SF, which locks up gas into stars;
\item
stellar ejecta which shed back enriched material to the interstellar medium
(ISM);
\item
infall of primordial protogalactic gas;
\item
gas exchange with the neighbouring rings because of radial flows.
\end{enumerate}
The set of equations driving the chemical evolution of the $k$-th shell is:
\begin{equation}
\label{dGi/dt_radf}
\begin{array}{l l}
\frac{d}{dt} G_i(r_k,t) = & - X_i(r_k,t) \Psi(r_k,t) \,+ \\
 & \\
 & +\, \int_{M_l}^{M_u} \Psi(r_k,t-\tau_M)\,R_i(M) \Phi(M) dM \,+ \\
 & \\
& + \, \left[\frac{d}{dt} G_i(r_k,t) \right]_{inf} \,+\, \\
 & \\
 & +\, \left[\frac{d}{dt} G_i(r_k,t) \right]_{rf}
\end{array}
\end{equation}
where the various symbols are defined here below.

Primordial gas is accreted at an exponentially decreasing rate with time-scale 
$\tau$:
%
\begin{equation}
\label{eqinfall}
\dot{\sigma}_{inf}(r_k,t) = A(r_k) \, e^{-\frac{t}{\tau(r_k)}}
\end{equation}
$A(r_k)$ is obtained by imposing that the integrated contribution of infall
up to the present Galactic age $t_G=$15~Gyr, corresponds 
to an assumed exponential profile $\sigma_A(r_k)$:
\begin{equation}
\label{Arprofile}
 A(r_k) = \frac{\sigma_A(r_k)}{\tau(r_k) (1-e^{-t_G/\tau(r_k)})} = 
\frac{\sigma_A(r_{\odot}) \, e^{-\frac{r_k-r_{\odot}}{r_d}}}
{\tau(r_k) (1-e^{-t_G/\tau(r_k)})}
\end{equation}
Indicating with $\sigma_g(r_k,t)$ the surface gas density, we define the gas 
fraction: 
\begin{equation}
\label{Gkdefinition}
G(r_k,t) = \frac{\sigma_g(r_k,t)}{\sigma_A(r_k)}
\end{equation}
and the normalized surface gas density for each chemical species $i$:
\begin{equation}
\label{Gikdefinition}
G_i(r_k,t) = X_i(r_k,t) \, G(r_k,t)
\end{equation}
where $X_i$ is the fractionary abundance by mass of $i$. 

The 1$^{st}$ term on the right-hand side of Eq.~(\ref{dGi/dt_radf}) represents
the depletion of species $i$ from the ISM due to star formation;
see PC99 for the various options concerning the SF rate, $\Psi(r,t)$.
The 2$^{nd}$ term is the amount of species $i$ ejected back to the ISM 
by dying stars; the returned fractions $R_i(M)$ are calculated on the base 
of the detailed stellar yields from PCB98 and keep track of finite stellar 
lifetimes (no instantaneous recycling approximation IRA). The 3$^{rd}$ term
is the contribution of infall, while the 4$^{th}$ term describes the effect
of radial flows. 
Full details on the first three terms can be found in the original static
model by PCB98 and PC99.
The novelty in Eq.~(\ref{dGi/dt_radf}) is the radial flow term, which we 
develop here below. We will adopt the simplified notation 
$\sigma_{g k} \equiv \sigma_g(r_k,t)$ and the like.

\begin{figure}[t]
\centerline{\psfig{file=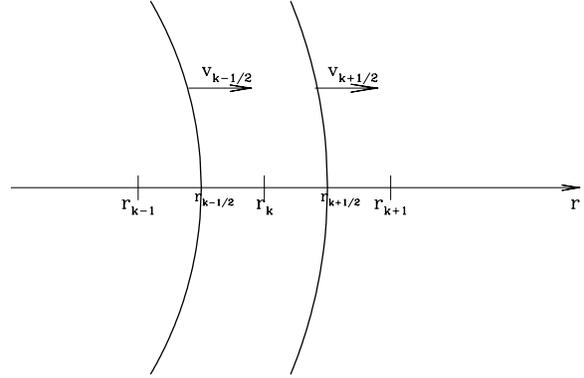,angle=-90,width=8truecm}}
\caption{Scheme of the gas flow through the $k$-th model shell}
\label{shellfig}
\end{figure}

Let the $k$-th shell be defined by the galactocentric radius $r_k$, its inner
and outer edge being labelled as $r_{k-\frac{1}{2}}$ and $r_{k+\frac{1}{2}}$. 
Through these edges, gas flows 
with velocity $v_{k-\frac{1}{2}}$ and $v_{k+\frac{1}{2}}$, respectively 
(Fig.~\ref{shellfig}). Flow velocities are taken positive outward;
the case of inflow is correspondingly 
described by negative velocities. 
%
%
Radial flows through the borders, with a flux $F(r)$, contribute to alter 
the gas surface density in the $k$-th shell according to:
\begin{equation}
\label{dsigmarf1}
\left[ \frac{d \sigma_{g k}}{d t} \right]_{rf} = 
   - \frac{1}{\pi \left( r^2_{k+\frac{1}{2}} - r^2_{k-\frac{1}{2}} \right) }
   \left[ F(r_{k+\frac{1}{2}}) - F(r_{k-\frac{1}{2}}) \right]
\end{equation}
The gas flux at $r_{k+\frac{1}{2}}$ can be written:
\begin{small}
\begin{equation}
\label{flux3}
F(r_{k+\frac{1}{2}}) = 2 \pi r_{k+\frac{1}{2}} \, v_{k+\frac{1}{2}} \left[
\chi(v_{k+\frac{1}{2}}) \, \sigma_{g k} + \chi(-v_{k+\frac{1}{2}}) \,
\sigma_{g (k+1)} \right]
\end{equation}
\end{small}
where $\chi(x)$ is the step function: {\mbox{$\chi(x)=1$ or 0}} for 
{\mbox{$x >$ or $\leq 0$}}, respectively.
Eq.~(\ref{flux3}) is a sort of ``upwind approximation'' for the advection
term to be included in the model equations (e.g.\
Press \etal 1986), describing either inflow or outflow depending on the sign
of $v_{k+\frac{1}{2}}$. An analogous expression holds for 
$F(r_{k-\frac{1}{2}})$.

Let's take the inner edge $r_{k-\frac{1}{2}}$
at the midpoint between $r_{k-1}$ and $r_k$,
and similarly for $r_{k+\frac{1}{2}}$ (Fig.~\ref{shellfig}).
Writing Eq.~(\ref{dsigmarf1}) separately for each chemical species $i$,
in terms of the $G_i$'s we obtain
the radial flow term of Eq.~(\ref{dGi/dt_radf}) as:
\begin{equation}
\label{dGirf}
\begin{array}{l l}
\left[ \frac{d}{dt} G_i(r_k,t) \right]_{rf} = & \alpha_k \, G_i(r_{k-1},t) 
\,-\, \beta_k \, G_i(r_k,t) \,+ \\
 & +\, \gamma_k \, G_i(r_{k+1},t)
\end{array}
\end{equation}
where:
\begin{equation}
\label{coeffradf}
\begin{array}{l l}
\alpha_k = & \frac{2}{r_k + \frac {r_{k-1} + r_{k+1}}{2}} 
	     \left[ \chi(v_{k-\frac{1}{2}}) \, v_{k-\frac{1}{2}} \, 
	     \frac{r_{k-1}+r_k}{r_{k+1}-r_{k-1}} \right] \, 
	     \frac{\sigma_{A (k-1)}}{\sigma_{A k}} \\
 & \\
\beta_k = & - \, \frac{2}{r_k + \frac {r_{k-1} + r_{k+1}}{2}} \times \\
	    \multicolumn{2}{l}{
	    \times \left[ \chi(-v_{k-\frac{1}{2}}) v_{k-\frac{1}{2}}
	    \frac{r_{k-1}+r_k}{r_{k+1}+r_{k-1}} - \chi(v_{k+\frac{1}{2}})
	    v_{k+\frac{1}{2}} \frac{r_k+r_{k+1}}{r_{k+1}-r_{k-1}} \right] } \\
 & \\
\gamma_k = & - \frac{2}{r_k + \frac {r_{k-1} + r_{k+1}}{2}} 
	     \left[ \chi(-v_{k+\frac{1}{2}}) v_{k+\frac{1}{2}}
	     \frac{r_k+r_{k+1}}{r_{k+1}-r_{k-1}} \right] 
	     \frac{\sigma_{A (k+1)}}{\sigma_{A k}}
\end{array}
\end{equation}
The terms on the right-hand side of Eq.~(\ref{dGirf}) 
evidence the contribution of the 3 contiguous shells involved: 
the first term represents the gas being gained in shell $k$ from $k$--1, 
the second term is the gas being lost from $k$ to $k$--1 and $k$+1, 
and the third term is the gas being gained in $k$ from $k$+1.
The coefficients~(\ref{coeffradf}) are 
all $\geq 0$ and depend only on the shell $k$, not on the chemical species $i$
considered. If the velocity pattern is constant in time, $\alpha_k$,
$\beta_k$ and $\gamma_k$ are also constant in time.

Notice that in the case of static models 
the final surface mass density is completely determined by the
assumed accretion profile, namely 
{\mbox{$\sigma(r_k,t_G) \equiv \sigma_A(r_k)$}}. Therefore, in static models 
the radial profile for accretion can be directly chosen so as to match the 
observed present--day surface density in the Disc (see PCB98 and PC99). 
The inclusion of the term of radial gas flows alters the expected final density
profile
and {\mbox{$\sigma(r_k,t_G) \neq \sigma_A(r_k)$}}.
Hence, $\sigma(r,t_G)$ cannot be assumed in advance and is only known 
{\it a posteriori} (see \S\ref{radfloweffects}); 
at the end of each simulation we need to check how much radial 
flows have altered the actual density profile $\sigma(r_k,t_G)$ with respect 
to the pure accretion profile $\sigma_A(r_k)$. With the slow 
flow speeds considered ($v \lsim 1$~km~sec$^{-1}$), the two profiles will not 
be too dissimilar anyways.


\subsection{Boundary conditions}

Eq.~(\ref{dGirf}) needs to be slightly modified in the case of the innermost
and the outermost shell, since the shell $k$--1 or $k$+1 
are not defined in these two respective cases.


\subsubsection{The innermost shell}
\label{shell1}

Our disc models will extend down to where the Bulge becomes 
the dominating Galactic component ($r_1=2.5$~kpc). As to the innermost edge, 
we assume that the first 
shell is symmetric with respect to $r_1$:
\[ r_{\frac{1}{2}} = \frac{3 r_1 - r_2}{2} \]
and that {\mbox{$v_{\frac{1}{2}} \leq 0$}} always, since
we cannot account for outflows from still inner shells, not included in the 
model. For {\mbox{$k=1$}}, Eq.~(\ref{dGirf}) then becomes:
\begin{equation}
\label{dGirf1}
\left[ \frac{d}{dt} G_i(r_1,t) \right]_{rf} = - \beta_1 G_i(r_1,t) \, + \, 
\gamma_1 G_i(r_2,t)
\end{equation}
with:
\[ \beta_1 = -\frac{1}{2 r_1} \left[ v_{\frac{1}{2}}\,\frac{3 r_1-r_2}{r_2-r_1}
\,-\, \chi(v_{\frac{3}{2}}) \, v_{\frac{3}{2}} \, \frac{r_1+r_2}{r_2-r_1} 
\right] \]
\begin{equation}
\label{coeffradf1}
\gamma_1 = - \chi(-v_{\frac{3}{2}}) \, v_{\frac{3}{2}} \, \frac{1}{2 r_1} \,
\frac{r_1+r_2}{r_2-r_1}\, \frac{\sigma_{A 2}}{\sigma_{A 1}}~~~~~~~~~~
\end{equation}


\subsubsection{Boundary conditions at the disc edge}
\label{boundary}

As to the outermost shell ($k=N$), 
we need a boundary condition for the gas inflowing from the outer disc. 
We assume
a SF cut-off in the outer disc, while the gaseous layer extends much further.
In fact, in external spirals HI discs are observed to extend much beyond the
optical disc, out to 2 or even 3 optical radii. A threshold preventing 
SF beyond a certain radius is expected from gravitational stability
in fluid discs (Toomre 1964, Quirk 1972) and has observational support as well
(Kennicutt 1989). For the Galactic Disc we assume no SF 
beyond the last shell at $r_N = 20$~kpc, which is the empirical limit for the
optical disc and for HII regions and bright blue stars tracing active SF.
Gas can though flow in from the outer disc;
extended gas discs might actually provide a much larger gas reservoir 
for star--forming spirals than vertical infall, at least at the present time 
when the gravitational settling of the protogalactic cloud is basically over.

If no SF occurs in the outer disc, the evolution of the gas (and total) surface
density can be expressed as:
\begin{equation}
\label{radfnoSF}
\frac{\partial \sigma}{\partial t} (r,t) = A(r) \, e^{-\frac{t}{\tau(r)}} 
\,-\, \frac{1}{r} \, \frac{\partial}{\partial r} (r v \sigma)~~~~~~~
\forall \, r>r_{N+\frac{1}{2}}
\end{equation}
(e.g.\ Lacey \& Fall 1985, their model equation with the SF term dropped).
Here, with no SF, $\sigma \equiv \sigma_g$ and abundances always
remain the primordial ones ($X_{i, inf}$).
Let's assume the following simplifying conditions for the outer disc:
\begin{enumerate}
\item
the infall time-scale is uniform: 
\[ \tau(r) \equiv \tau(r_N)~~~~~~~~~~~~~~~~~~~~~
\forall \, r>r_{N+\frac{1}{2}} \]
\item
the inflow velocity is uniform and constant: 
\[ v(r,t) \equiv v_{N+\frac{1}{2}}~~~~~~~~~~~~~~~~~~~
\forall \, r>r_{N+\frac{1}{2}}, ~~\forall \, t\]
\item
the infall profile 
is flat:
\[ A(r) \equiv A_{ext} ~~~~~~~~~~~~~~~~~~~~~~~\forall \, r>r_{N+\frac{1}{2}} \]
in accordance with observed extended gas discs in spirals, showing a much 
longer scale-length than the stellar component.
\end{enumerate}
With these assumptions, 
Eq.~(\ref{radfnoSF}) becomes:
\begin{equation}
\label{bordereq}
\frac{\partial \sigma}{\partial t}\,+\,v\,\frac{\partial \sigma}{\partial r}
 = A \, e^{-\frac{t}{\tau}} \,-\, \frac{v}{r} \,\sigma
\end{equation}
where we indicate $\tau \equiv \tau(r_N)$, $v \equiv v_{N+\frac{1}{2}}$ and
$A \equiv A_{ext}$ to alleviate the notation.
Eq.~(\ref{bordereq}) has a straightforward analytical solution 
(Appendix~B):
\begin{equation}
\label{borderconditionTrf}
\begin{array}{l l}
\sigma(r,t) = & A \, \tau \, \times \\
 & \\
\multicolumn{2}{r}{ \times \left[ \left( 1 - e^{-\frac{t}{\tau}} \right)
+ \frac{v}{r} \left( \tau \left( e^{-\frac{T_{rf}}{\tau}} - 
e^{-\frac{t}{\tau}} \right) - (t - T_{rf}) \right) \right] }
\end{array}
\end{equation}
where $T_{rf} \geq 0$ is the time when radial inflows are assumed to activate.
Eq.~(\ref{borderconditionTrf}) is our
boundary condition at the outermost edge.

Notice that (\ref{borderconditionTrf})
is the solution of (\ref{bordereq}) 
in the idealized case of an infinite, flat gas layer extending boundless 
to any $r > r_N$ (see also Appendix~B).
Of course, this
does not correspond to gaseous discs surrounding real spirals; but since
we will consider only slow inflow velocities ($v \lsim 1$~km~sec$^{-1}$), with
typical values of $r_N=20$~kpc and $t_G=15$~Gyr, the gas actually drifting into
the model disc shells will be just the gas originally accreted within 
$r \sim 35$~kpc. Therefore, the boundary condition~(\ref{borderconditionTrf})
remains valid as long as the gas layer stretches out to $\sim 35$~kpc,
a very plausible assumption since observed gaseous discs extend over 
a few tens or even {\mbox{$\sim 100$~kpc}}.


\subsubsection{The outermost shell}
\label{shellN}

We take a reference external radius $r_{ext} > r_N$ in the outer disc where 
the (total and gas) surface density 
$\sigma(r_{ext},t) \equiv \sigma_{ext}(t)$ is given by the boundary condition
(\ref{borderconditionTrf}); typically, $r_N =20$~kpc and $r_{ext} \sim 21$~kpc.
We take the outer edge of the shell at the midpoint:
\[ r_{N+\frac{1}{2}} = \frac{r_N + r_{ext}}{2} \]
and replace
\[ X_{i (k+1)} \, \sigma_{g (k+1)} \longrightarrow X_{i,inf} \, \sigma_{ext} \]
in Eqs.~(\ref{dsigmarf1}) and~(\ref{flux3}), since 
the primordial abundances $X_{i,inf}$ remain unaltered in the outer disc, 
in the absence of SF.
We thus write the radial flow term for the $N$-th shell as: 
\begin{equation}
\label{dGirfN}
\begin{array}{l l}
\left[ \frac{d}{dt} G_i(r_N,t) \right]_{rf} = & \alpha_N \, G_i(r_{N-1},t)
\,-\, \beta_N \, G_i(r_N,t) \,+ \\
                                              & +\, \omega_i(t)
\end{array}
\end{equation}
where:
\begin{equation}
\label{coeffradfN}
\begin{array}{l l}
\omega_i = & - X_{i,inf} \,\,\, \chi(-v_{N+\frac{1}{2}}) \,\, v_{N+\frac{1}{2}}
\, \times \\
 & \times \, 
\frac{4}{r_{N-1} + 2 r_N + r_{ext}}\,\,\frac{r_N+r_{ext}}{r_{ext}-r_{N-1}}
\,\, \frac{\sigma_{ext}(t)}{\sigma_{A N}}
\end{array}
\end{equation}
%


\subsection{The numerical solution}
\label{numericalradf}

Using~(\ref{dGirf}), (\ref{dGirf1}) and (\ref{dGirfN}), the basic set of 
equations~(\ref{dGi/dt_radf}) can be written as:
\[ \left\{ \begin{array}{l l}
\frac{d}{dt} \, G_i(r_1,t) = & \vartheta_1(t) \, G_i(r_1,t) +
			       \gamma_1 G_i(r_2,t) + W_i(r_1,t) \\
 & \\
\frac{d}{dt} \, G_i(r_k,t) = & \alpha_k G_i(r_{k-1},t) + 
			       \vartheta_k(t) \, G_i(r_k,t) + \\
 & + \gamma_k G_i(r_k,t) + W_i(r_k,t) \\
\multicolumn{2}{r}{k=2,...N-1} \\
 & \\
\frac{d}{dt} \, G_i(r_N,t) = & \alpha_N G_i(r_{N-1},t) + 
			       \vartheta_N(t) \, G_i(r_N,t) + \\
			     & + W_i(r_N,t) + \omega_i(t)
\end{array} \right. \]
where we have introduced: 
\[ \begin{array}{l c l}
\vartheta_k(t) & \equiv & - \left( \eta(r_k,t) + \beta_k \right) \leq 0 \\
 & \\
\eta(r_k,t) & \equiv & \frac{\Psi}{G}(r_k,t) \\
 & \\
\medskip
W_i(r_k,t) & \equiv & \int_{M_l}^{M_u} \Psi(r_k,t-\tau_M) \, R_i(M) \Phi(M) dM 
\, + \\
           &        & +\, \left[\frac{d}{dt} G_i(r_k,t) \right]_{inf}
\end{array} \]
We refer to PCB98 for further details on the quantities 
$\eta$ and $W_i$, appearing also in the original static model.
Neglecting, for the time being, that the $\eta$'s and the $W_i$'s contain
the $G_i$'s themselves, we are dealing with a linear, first order, 
non homogeneous system of
differential equations with non constant coefficients, of the kind:
\begin{equation}
\label{systemradf}
\frac{d \vec{G}_i}{dt} \,=\,{\cal  A}(t) \, \vec{G}_i(t) \,+\, \vec{W}_i(t)
\end{equation}
There is a system~(\ref{systemradf}) for each chemical species $i$, but the 
matrix of the coefficients ${\cal A}(t)$ is independent of $i$.

We solve the system by the same numerical method used for the original
equation of the static model --- see Talbot \& Arnett (1971) and PCB98
for details. We just need to extend the method to the present 
multi--dimensional case~(\ref{systemradf}). 
If we consider the evolution of the $G_i$'s over a short enough timestep 
$t_1-t_0 = \Delta t$, the various quantities $\eta(r_k,t)$,
$\vartheta_k(t)$ and $W_i(r_k,t)$ will remain roughly constant within 
$\Delta t$; similarly to the method for the static model (see PCB98),
within $\Delta t$ we approximate them with the values $\overline{\eta}_k$, 
$\overline{\vartheta}_k$ and $\overline{W}_i(r_k)$ they assume at the midstep 
$t_{\frac{1}{2}} = t_0 + \frac{1}{2} \Delta t$. Over $\Delta t$,
(\ref{systemradf}) can then be considered a system with constant coefficients 
${\cal A} \equiv {\cal A}(t_{\frac{1}{2}})$, and $\vec{W}_i (t)$ becomes a 
constant vector, which allows for the analytical solution:
\begin{equation}
\label{solsystemradf2}
\vec{G}_i (t_1) = e^{\Delta t\, {\cal A}} \, \vec{G}_i (t_0) + \left[
\int_{t_0}^{t_1} e^{(t_1-t) {\cal A}} \, dt \right] \,\, \vec{W}_i
\end{equation}
where 
$e^{t {\cal A}}$ indicates the matrix:
\begin{equation}
\label{e^tAformula}
e^{t {\cal A}} = \pmatrix{ e^{\lambda_1 t} \vec{u}_1 & 
\ldots & e^{\lambda_N t} \vec{u}_N } 
\pmatrix{ \vec{u}_1 & 
\ldots & \vec{u}_N }^{-1}
\end{equation}
with $\lambda_k$ the eigenvalues of ${\cal A}$ and 
$\vec{u}_k$ the corresponding eigenvectors.
The matrix $e^{t {\cal A}}$ and the explicit expression of the 
solution~(\ref{solsystemradf2}) for the $G_i$'s are 
calculated in Appendix~A,
resulting in:
\begin{equation}
\label{solradfN}
\begin{array}{l l}
G_i(r_k,t_1) = & \alpha_k \, \times \\ 
 	\multicolumn{2}{r}{ \times \, 
	 \frac{e^{-(\overline{\eta}_k+\beta_k) \Delta t}-
	e^{-(\overline{\eta}_{k-1}+\beta_{k-1}) \Delta t}}
	{(\overline{\eta}_{k-1}+\beta_{k-1})-(\overline{\eta}_k+\beta_k)} \, 
	G_i(r_{k-1},t_0) \,+} \\
 &	+\, e^{-(\overline{\eta}_k+\beta_k) \Delta t} \, G_i(r_k,t_0) \,+\\
\multicolumn{2}{r}{
	+\, \gamma_k \,\frac{e^{-(\overline{\eta}_{k+1}+\beta_{k+1}) \Delta t}-
	e^{-(\overline{\eta}_k+\beta_k) \Delta t}}{(\overline{\eta}_k+\beta_k)-
	(\overline{\eta}_{k+1}+\beta_{k+1})} \, G_i(r_{k+1},t_0) \,+ }\\
 &	+\, \alpha_k\,\frac{\overline{W}_i(r_{k-1})}
	{(\overline{\eta}_{k-1}+\beta_{k-1})-
	(\overline{\eta}_k+\beta_k)} \, \times \\
	\multicolumn{2}{r}{ \times \left(
  \frac{1-e^{-(\overline{\eta}_k+\beta_k) \Delta t}}{\overline{\eta}_k+\beta_k}
	- \frac{1-e^{-(\overline{\eta}_{k-1}+\beta_{k-1}) \Delta t}}
	{\overline{\eta}_{k-1}+\beta_{k-1}} \right) \,+ }\\
 & +\, \overline{W}_i(r_k)\, \frac{1-e^{-(\overline{\eta}_k+\beta_k) \Delta t}}
	{\overline{\eta}_k+\beta_k} \,+ \\
 &	+\, \gamma_k\,\frac{\overline{W}_i(r_{k+1})}
	{(\overline{\eta}_k+\beta_k)-(\overline{\eta}_{k+1}+\beta_{k+1})}
	\, \times \\
 &	\times
	 \left(\frac{1-e^{-(\overline{\eta}_{k+1}+\beta_{k+1}) \Delta t}}
	{\overline{\eta}_{k+1}+\beta_{k+1}} - 
	\frac{1-e^{-(\overline{\eta}_k+\beta_k) \Delta t}}
	{\overline{\eta}_k+\beta_k} \right) \\  
\end{array}
\end{equation}
where we intend $\alpha_1 \equiv 0$, $\gamma_N \equiv 0$, 
and for the outermost shell one should replace: 
\[ \overline{W}_i(r_N) \longrightarrow 
\overline{W}_i(r_N) + \overline{\omega}_i =  \overline{W}_i(r_N) +
\frac{1}{\Delta t} \, \int_{t_0}^{t_1} \omega_i(t) \, dt \]
$\omega_i$ being defined by~(\ref{coeffradfN}).

Similarly to the case of an isolated shell (see PCB98) the 
system~(\ref{solradfN}) 
does not provide the final solution for $G_i(r_k,t_1)$, since the 
$\overline{\eta}_k$'s and the $\overline{W}_i$'s on the right--hand side
actually depend on the  $G_i$'s;
it just represents 
a set of implicit non--linear expressions in $G_i(r_k,t_1)$.
As in the original static model, we can neglect the dependence of 
$\overline{W}_i(r_k)$ on the  $G_i$'s and consider only that of 
$\overline{\eta}_k$ (see PCB98 for a detailed discussion). 
We must finally find the roots of the system~(\ref{solradfN}) by applying the 
Newton-Raphson method, generalized to many dimensions (cfr.\ Press et~al., 
1986). 
Such a system holds for each of the chemical species $i$ considered, so at 
each iteration we actually need to solve as many systems as the species 
included in the model.
Full details on the mathematical development of the model and its numerical 
solution can be found in the appendices and in Portinari (1998).

We tested the code against suitable analytical counterparts
and obtained the following conditions for model consistency (Appendix~B).
\begin{enumerate}
\item
Rather small timesteps are needed for the numerical model to keep stable;
the required timesteps get smaller and smaller the higher the flow velocities 
considered, and the thinner the shells.
\item
To describe gas flows in a disc with an exponential density profile, the shells
should be equispaced in the logarithmic, rather than linear, scale
(so that they roughly have the same mass, rather than the same width).
\end{enumerate}
We modelled the Galactic Disc using 35 shells
from 2.5 to 20 kpc, equally spaced in the logarithmic scale, their width
ranging from $\sim 0.2$~kpc for the inner shells to $\sim 1$~kpc for the 
outermost ones. With such a grid spacing, and velocities up to 
{\mbox{$\sim 1$~km~sec~$^{-1}$}}, 
suitable timesteps are of $10^{-4}$~Gyr (Appendix~B; see also 
Thon \& Meusinger 1998). This means that roughly {\mbox{$1.5 \times 10^5$}} 
timesteps, times 35 shells, are needed to complete each model, which would
translate in excessive computational times. This drawback was avoided 
by separating the time-scales in the code.
\begin{enumerate}
\item
The timestep $\Delta t$ used to update the ``chemical'' variables 
($\eta$, $W_i$, etc.) is 
the minimum
among: $\Delta t_1$ which guarantees that the relative variations 
of the $G_i$'s are lower than a fixed $\epsilon$; $\Delta t_2$ which
guarantees that the total surface mass density $\sigma(t)$ increases 
by no more than 5\%; $\Delta t_3$ which is twice the previous timestep of the
model, to speed up the computation when possible;
$\Delta t_4$ which guarantees 
the Courant condition {\mbox{$\Delta t < v \, \Delta r$}}, 
indispensable for the stability of a numerical algorithm describing
flows. So, $\Delta t$ is basically set by the requirement that
the chemical quantities do not vary too much within it, and it can get
relatively large (up to 0.2~Gyr), especially at late ages when the various 
chemical variables evolve slowly.
\item
It is only the numerical solution~(\ref{solradfN}) which needs very short 
timesteps to keep stable. Therefore, once the chemical variables are upgraded,
the main timestep $\Delta t$ is subdivided in much shorter timesteps 
{\mbox {$\delta t = 10^{-4}$~Gyr}}, upon which the solution~(\ref{solradfN})
and its Newton-Raphson iteration
are successively
applied to cover the whole $\Delta t$. Only then a new upgrade of all the
$\overline{\eta}_k$'s and $\overline{W}_i$'s is performed.
\end{enumerate}
This trick keeps the code 
roughly as fast as if it would evolve with a single 
time-scale $\Delta t$, and yet it gives the
same results as the ``slow'' version in which all quantities are upgraded at
each $\delta t = 10^{-4}$~Gyr.

\begin{table}[hb]
\caption{Parameter values and resulting metallicity gradients for models
{\sf S15RF}, {\sf O10RF} and {\sf DRRF}}
\label{modelRFtab}
\begin{tabular}{l|l|l|l|l|c|l}
\hline
 & & & & & & \\
 model & $r_d$ & $\nu$ & $\zeta$ & $\tau$ & $v(r)$ & $\frac{d[O/H]}{dr}$ \\
 & & & & & & \\
\hline
\multicolumn{7}{c}{} \\
\hline
{\sf S15a} & 4 & 0.35 & 0.2  & 3 & 0 & --0.03 \\
\hline
{\sf S15RFa} & 4 & 0.35 & 0.2  & 3 & --1 & --0.053 \\
 & & & & & & \\
{\sf S15RFb} & 5 & 0.4  & 0.32 & 3 & --1 & --0.047 \\
\hline
{\sf S15RFc} & 4 & 0.35 & 0.2  & 3 & $\left\{ \begin{array}{c}
						-1 \\
				     v_{N+\frac{1}{2}}=0
				     \end{array} \right.$ & --0.083 \\
 & & & & & & \\
{\sf S15RFd} & 7 & 0.35 & 0.35 & 3 & $\left\{ \begin{array}{c}
						-1 \\
				     v_{N+\frac{1}{2}}=0
				     \end{array} \right.$ & --0.073 \\
\hline
{\sf S15RFe} & 6 & 0.37 & 0.25 & 3 & Fig.~\ref{velS15bestfit} & --0.063 \\
\hline
\multicolumn{7}{c}{} \\
\hline
{\sf O10a} & 4   & 0.19 & 0.2  & 3 & 0 & --0.03 \\
\hline
{\sf O10RFe} & 7   & 0.2  & 0.25 & 3 & Fig.~\ref{velS15bestfit} & --0.07 \\
\hline
\multicolumn{7}{c}{} \\
\hline
{\sf DRa} & 4   & 0.42 & 0.2   & 3 & 0 & 
--0.07 {\scriptsize ($r > r_{\odot}$)} \\
	&	&    &      &   &   & ~flat~~{\scriptsize ($r < r_{\odot}$)} \\
\hline
{\sf DRRFe} & 4   & 0.5  & 0.27 & 3 & Fig.~\ref{velTA15bestfit} & --0.059 \\
\hline
\end{tabular}
\end{table}

\begin{figure*}[t]
\centerline{\psfig{file=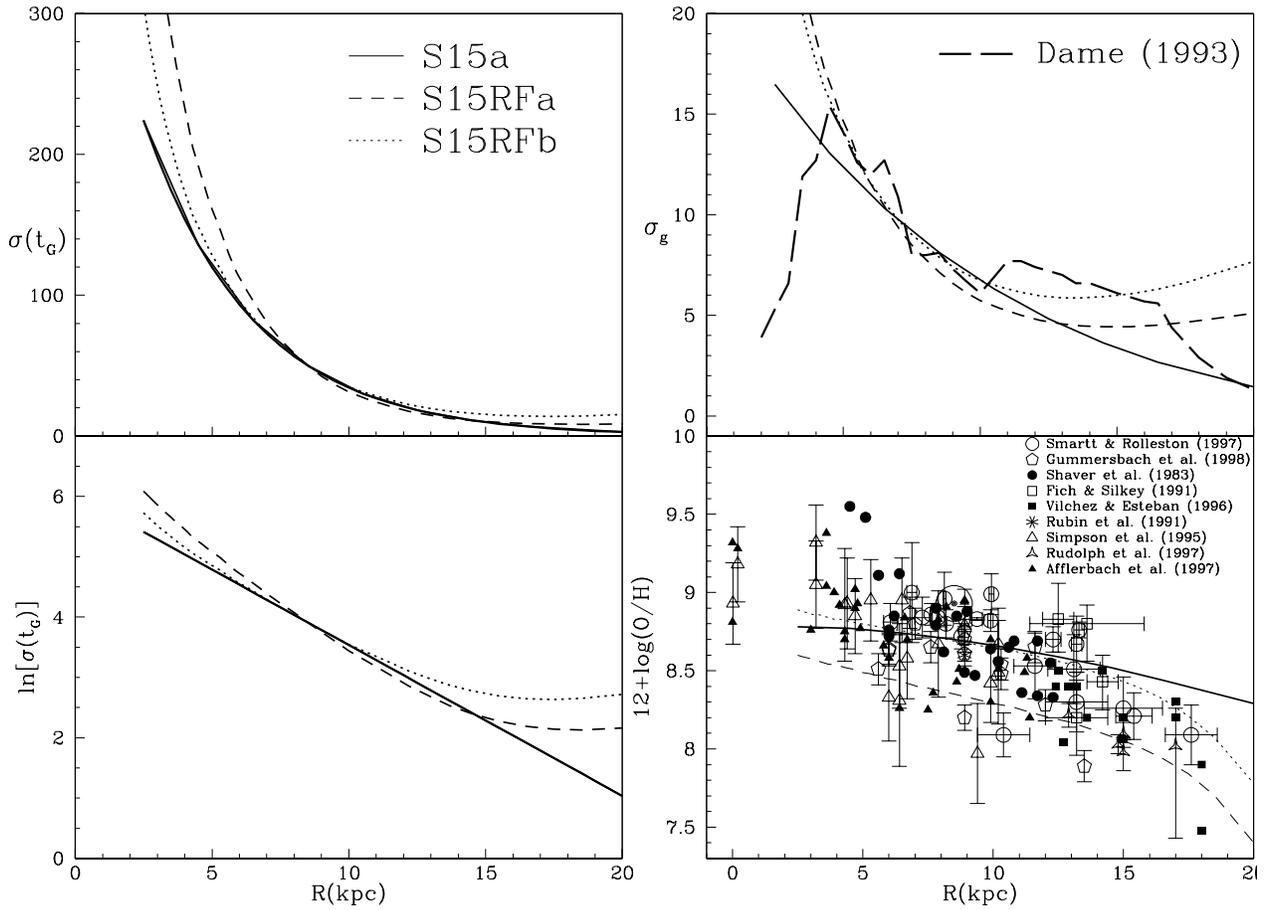,angle=-90,width=17truecm}}
\caption{{\it Left panels}: final profiles of the total surface density 
from models {\sf S15RF} with inflow from the outer disc, in linear (upper 
panel) or logarithmic (lower panel) plot; the reference observed profile 
of models with scale length 4~kpc,
as adopted in models with no flows (PC99), is shown as a thick solid line.
{\it Upper right panel}: radial gas density profiles compared to the
observed one. {\it Lower right panel}: predicted radial metallicity profiles 
compared to the observational data (data and symbols as in PC99).}
\label{s15rf}
\end{figure*}


\section{Effects of radial inflows on the disc}
\label{radfloweffects}

After presenting our new model with radial flows, we investigate
their general effects on the chemical evolution of the disc. 
Here we first analyse how the predicted metallicity gradient and gas and total 
surface density profiles of a static model are
altered by a superimposed uniform gas inflow,
while in Section~\ref{bestfit} we will suitably combine radial flows 
with various SF laws to match the observed radial profile of the Disc.
The characteristics and parameters of the various models presented in 
this and in the next section are summarized in Table~\ref{modelRFtab}.

To gain a qualitative understanding of the effects of radial gas inflows on 
chemical evolution, we impose a uniform inflow pattern $v=-1$~km~sec$^{-1}$
upon a static chemical model
and compare the new outcome with the original static case. 
Radial flows over the disc are expected to be mainly 
inflows, with velocities from 0.1 to $\sim$1~km~sec$^{-1}$ 
(see \S\ref{previous}); imposing a uniform inflow
of 1~km~sec$^{-1}$ is therefore a sort of ``extreme case'',
considered here for the sake of qualitative analysis. Anyways, previous
studies in literature suggest that the effects of radial flows saturate
for much higher velocities (K\"oppen 1994). We will 
consider both the case of inflow from the outer gaseous disc and not 
({\mbox{$v_{N+\frac{1}{2}}=-1$}} or 0, respectively).

All models with radial flows are rescaled 
so that the final surface density at the Solar ring ($r_{\odot}=8.5$~kpc) 
corresponds to 50~\Msol~pc$^{-2}$. Namely, 
with radial flows $\sigma(r_{\odot},t_G) \neq \sigma_A(r_{\odot})$ 
and it cannot be imposed as an input datum (see \S\ref{discrete}), but the 
zero--point $A(r_{\odot})$ of the exponential accretion 
profile~(\ref{Arprofile})
\[ A(r) = A(r_{\odot}) \,\, e^{-\frac{r-r_{\odot}}{r_d}} \]
%
\begin{figure*}[t]
\centerline{\psfig{file=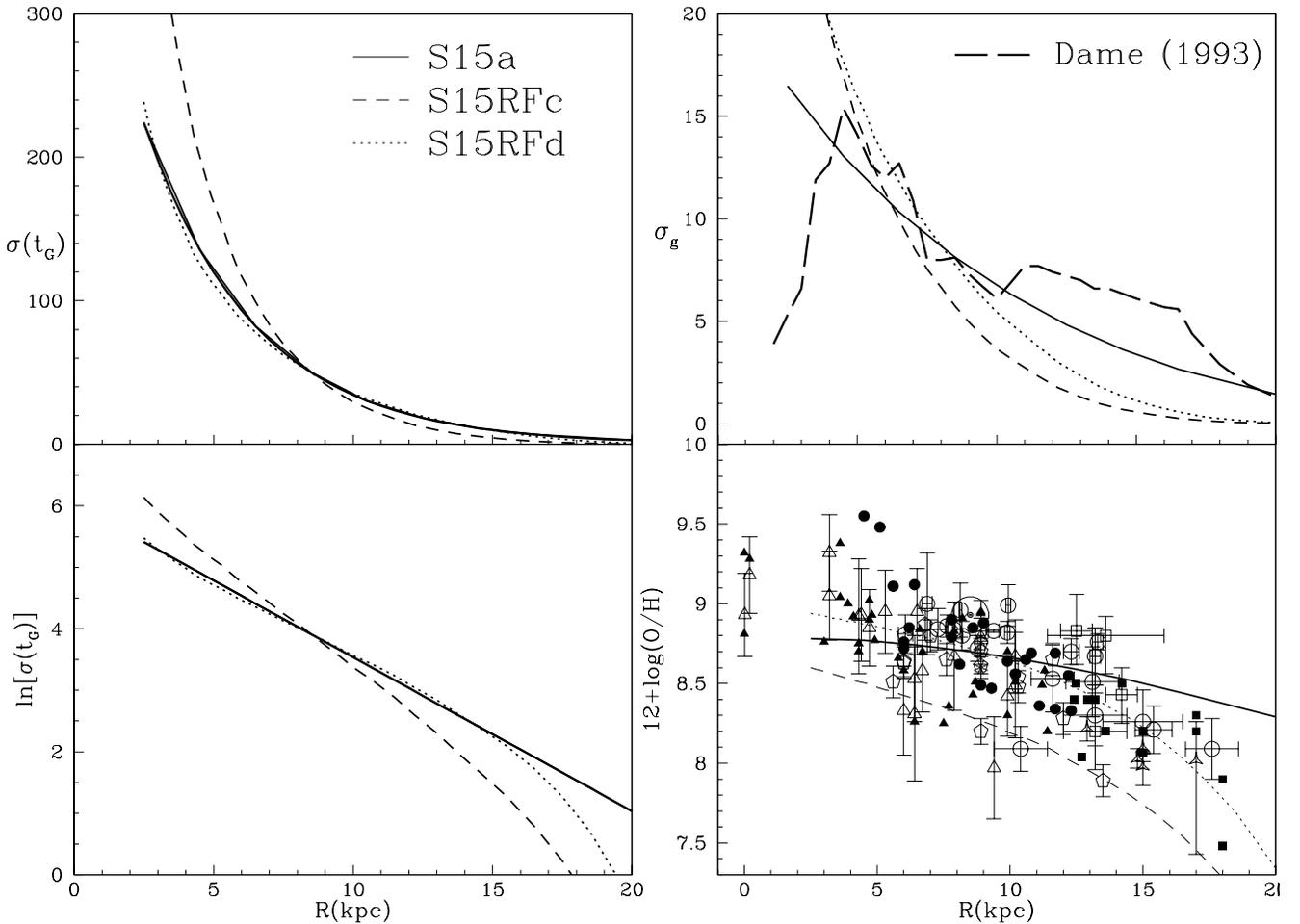,angle=-90,width=18truecm}}
\caption{Same as Fig.~\protect{\ref{s15rf}},
but for models {\sf S15RF} with no inflow from the outer disc edge.}
\label{s15rfnoext}
\end{figure*}
%
can be rescaled so that at the end of the simulation 
$\sigma(r_{\odot},t_G)=50$~\Msol~pc$^{-2}$.
This zero--point does not influence the profile, nor the chemical 
evolution, so it can always be rescaled {\it a posteriori}.

For our example, we take as the reference static case a model adopting
a Schmidt SF law with $\kappa=1.5$ (Kennicutt 1998) and a uniform infall 
time-scale of 3~Gyr (model {\sf S15a} of PC99, see also 
Table~\ref{modelRFtab}).


\subsection{Models with inflow from the outer disc}

In model {\sf S15RFa} a uniform radial inflow pattern with 
{\mbox{$v=-1$~km~sec$^{-1}$}}
and inflow from the outer disc is imposed upon the static model {\sf S15a},
with no further change in the model parameters (Table~\ref{modelRFtab}). 
Fig.~\ref{s15rf} shows the effects of inflow 
on the total surface density and  gas density distribution, 
and on the oxygen gradient 
(details on the observational data in the plots can be found in PC99).
By comparing the solid line and the dashed line,
we notice the following main effects. 
\begin{itemize}
\item
Since matter flows inward and accumulates toward the Galactic Centre, the 
density profile gets steeper in the inner parts, while in the outer parts 
it remains rather flat because gas is continuously poured in by the flat outer
gaseous disc. The gas density distribution shows a similar behaviour.
\item
The overall metallicity gets much lower because of the dilution by
primordial gas inflowing from the outer disc. The gradient becomes steeper,
especially in the outermost shells,
since the discontinuity in metallicity between the star--forming 
disc and the outer gaseous layer is smeared inward by radial inflows.
\end{itemize}
To compensate for the steepening of the density distribution induced by radial
inflows, we must adopt a shallower initial accretion profile; 
our simulations show that a scale length $r_d \sim5$~kpc for the infall 
profile reduces in the end to a density profile matching the desired scale 
length of $\sim$4~kpc at the Solar Neighbourhood. At the same
time, the chemical enrichment must get more efficient for the overall
metallicity to increase to the observed levels; the predicted metallicity 
is improved by adopting an IMF more
weighted towards massive stars, i.e.\ by increasing the ``IMF scaling 
fraction'' $\zeta$ (see PCB98 and PC99 for a description of our model 
parameters). Together with the SF efficiency $\nu$, $\zeta$ is re-calibrated
to match the observed gas surface density and metallicity at the Solar 
Neighbourhood (as for the models of PC99).
In this way we calibrate model {\sf S15RFb} with respect to the Solar 
Neighbourhood; see Table~\ref{modelRFtab} for details on the adopted 
parameters. Comparing now this re-calibrated model with radial flows (dotted
line) to the original model {\sf S15a}, the metallicity gradient is increased
with respect to the static case, but still a bit flat with respect to 
observations. The gas density distribution peaks in the inner regions, as 
expected, and remains quite high (much higher than observed) in the outer 
regions due to substantial replenishment from the outer disc.


\subsection{Models with no inflow from the outer disc}

Let's now consider the case when radial flows are limited within the star
forming disc and there is no inflow from the external gaseous disc 
($v_{N+\frac{1}{2}}=0$). If radial flows are mainly driven by shocks in spiral
arms, for instance, they might indeed occur only within the 
stellar disc, while the outer gas layer remains unperturbed.

Fig.~\ref{s15rfnoext} (left panels) shows how the 
final density 
profile becomes much steeper than the reference accretion profile 
(model {\sf S15RFc}, dashed line), as expected since matter is efficiently
drifting inward with no replenishment from the outer disc.
For the final density profile to match the observed one, we must assume a much
shallower initial accretion profile ($r_d \sim 6-7$~kpc). 
Then the resulting local density profile is close to the observed one, 
while in the outer parts the profile remains steeper (model {\sf S15RFd}, 
dotted line).

The gas density profile shows a similar behaviour, strongly peaked
toward the centre while dropping quickly (much more quickly than observed) 
outside the Solar ring (upper right panel).

The overall metallicity is reduced with respect to the original model
{\sf S15a} (lower right panel, model {\sf S15RFc})
and again we need to increase the IMF scaling fraction $\zeta$ to raise
the chemical enrichment to the observed values (model {\sf S15RFd}). 
The resulting gradient is roughly comparable to the observed one.


\subsection{Concluding remarks}

We can summarize the general effects of radial flows as follows.
\begin{itemize}
\item
Since matter flows inward and accumulates toward the Galactic Centre, the 
density profile in the inner parts gets steeper. A shallower intrinsic 
accretion profile is to be adopted, in order to recover the observed local
scale--length at the end of the simulation.
\item
In the outer parts, the profile declines more or less sharply, depending on
whether the inflow occurs only within the star--forming disc or there is also
substantial inflow from the outer, purely gaseous disc.
\item
The gas distribution shows a similar behaviour: it remains quite
flat in the outer regions if gas is poured in by the flat outer gas disc,
while it tends to drop sharply (more than observed) otherwise.
\item
The inner gas profile is very steep in the case of a Schmidt SF law (models
{\sf S15RF}) --- while with other SF laws with radially decaying efficiency
(see PC99 and \S\ref{bestfit}) the effect would be somewhat compensated by
a larger gas consumption by SF in the inner regions.
\item
The overall metallicity gets much lower because of the dilution by
gas inflowing from metal poor outer shells (and possibly
the primordial outer disc). A higher fraction of stars contributing to the
chemical enrichment is needed in the model to match the observed metallicity.
\item
The metallicity gradient tends to steepen, in agreement with results by other 
authors (see references in \S\ref{previous}).
\end{itemize}


\section{Some ``successful'' models}
\label{bestfit}

In \S\ref{radfloweffects}
we have illustrated the qualitative effects of radial inflows using,
for the sake of example, models with a Schmidt SF law. We will now
consider models with various SF laws (see PC99) and
tune the inflow velocity pattern, in each case, so as to match the
observational data on the radial profile of the Galactic Disc.
The various ``successful'' models presented here should not be taken
as detailed, unique recipes to reproduce the Disc. Rather, they are meant
as examples of how inclusion of radial flows in the chemical model
can improve the match with the data.


\subsection{Models with Schmidt law}

Let's first consider again the Schmidt SF law. 
Inspection of 
models {\sf S15RFb} and {\sf S15RFd} (with and
without radial inflows from the outer disc) suggests to proceed as follows.
%
\begin{figure}[hb]
\centerline{\psfig{file=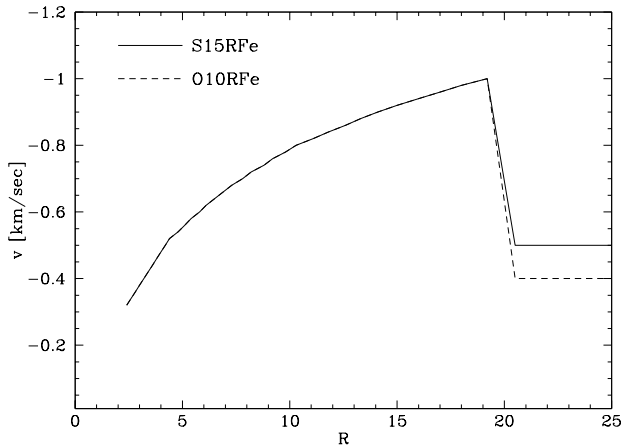,angle=-90,width=8.5truecm}} 
\caption{Inflow velocity pattern for the ``successful'' models {\sf S15RFe} 
and {\sf O10RFe}.}
\label{velS15bestfit}
\end{figure}
%
\begin{itemize}
\item
Some inflow from the outer gaseous layer is needed to reproduce the 
shallow decline of the gas distribution observed out of the Solar ring, though
the inflow speed from the outer disc should be slower than --1~km~sec$^{-1}$ 
otherwise the predicted gas density is too high in the outer parts (as in
model {\sf S15RFb}).
\item
Models with radial inflows can predict metallicity gradients close to the 
observed ones even with a Schmidt SF law, provided drift velocities within
the star--forming edge are relatively high (of the order of 
{\mbox{--1~km~sec$^{-1}$}}).
Inflow patterns decelerating inward are particularly efficient in 
building the metallicity gradient (G\"otz \& K\"oppen 1992).
\end{itemize}
%
\begin{figure}[t]
\centerline{\psfig{file=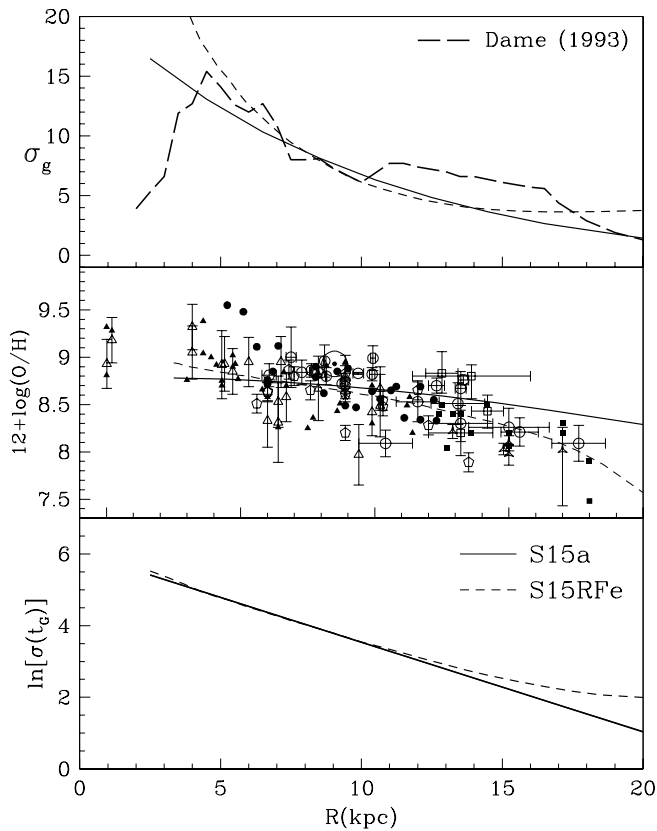,width=8.9truecm}}
\caption{A ``successful'' {\sf S15RFe} model compared to the case 
with no flows. Data and symbols as in Fig.~\protect{\ref{s15rf}}.}
\label{s15rfbestfit}
\end{figure}
%
An example of a drift velocity profile shaped according to these considerations
is shown in Fig.~\ref{velS15bestfit}; the corresponding model gives indeed 
a reasonable fit to the data (model {\sf S15RFe}, Fig.~\ref{s15rfbestfit}).
The gas profile keeps increasing 
inward, yet the model does not reproduce the peak of the molecular
ring, for which we need to include the peculiar radial flows induced 
by the Bar (see the discussion in PC99 and \S\ref{bar}). Model {\sf S15RFe} 
shows how radial inflows 
can in fact allow for metallicity gradients comparable with the observed
ones, even with a Schmidt SF law which would be excluded on the sole base of
static models (PC99).


\subsection{Models with spiral--triggered SF laws}

Let's consider now models adopting an Oort--type SF law, with
a SF efficiency inversely proportional to the galactocentric radius and a 
Schmidt--like exponent $\kappa=1.0$ (Kennicutt 1998, PC99). The 
static and the ``successful'' model with this SF law are model {\sf O10a} from 
PC99 and model {\sf O10RFe}, respectively (Table~\ref{modelRFtab}).

\begin{figure}[t]
\centerline{\psfig{file=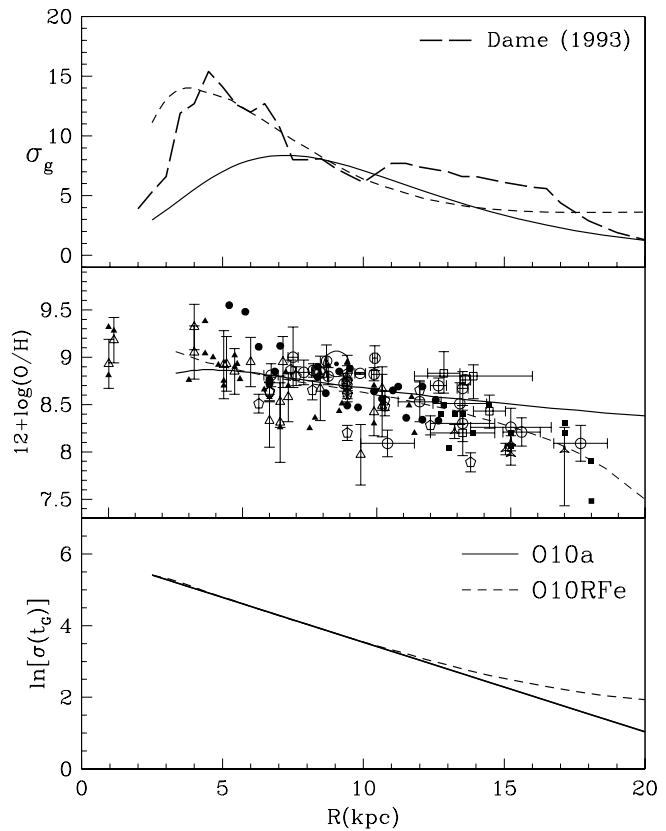,width=8.9truecm}}
\caption{A ``successful'' {\sf O10RF} model compared to the case 
with no flows.}
\label{ws10rfbestfit}
\end{figure}

As in the static case, 
(see PC99), these models
behave somewhat similarly to models
adopting the Schmidt SF law. Fig.~\ref{ws10rfbestfit} shows 
that a good fit is obtained with model {\sf O10RFe}, where we applied
the velocity  pattern of Fig.~\ref{velS15bestfit} with inflows becoming slower
inwar, very close to that used for model {\sf S15RFe}. Notice that 
in this case the higher SF efficiency in the inner region and the slowdown of 
inflows conspire to accumulate gas around $r=3$~kpc while consuming it at
inner radii, generating a peak in the predicted gas distribution which closely
reminds the observed molecular ring.


\subsection{Models with gravitational self--regulating SF laws}

We now investigate models adopting a self--regulating SF process driven by 
gravitational settling and feed--back from massive stars, implying a SF 
efficiency exponentially decaying outward in radius, such as the law by Dopita
\& Ryder (1994; see PC99). The reference static model
in this case is model {\sf DRa} of PC99, while the corresponding ``successful''
model with radial flows is {\sf DRRFe} (see Table~\ref{modelRFtab} for the 
relevant parameters).

\begin{figure}[t]
\centerline{\psfig{file=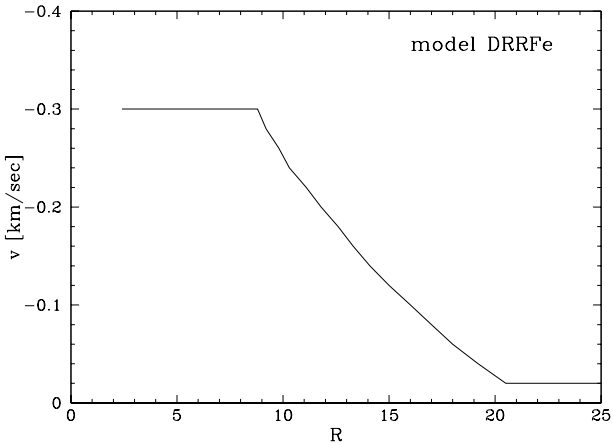,angle=-90,width=8.4truecm}}
\caption{Inflow velocity pattern for the ``successful'' model {\sf DRRF}}
\label{velTA15bestfit}
\end{figure}

\begin{figure}[t]
\centerline{\psfig{file=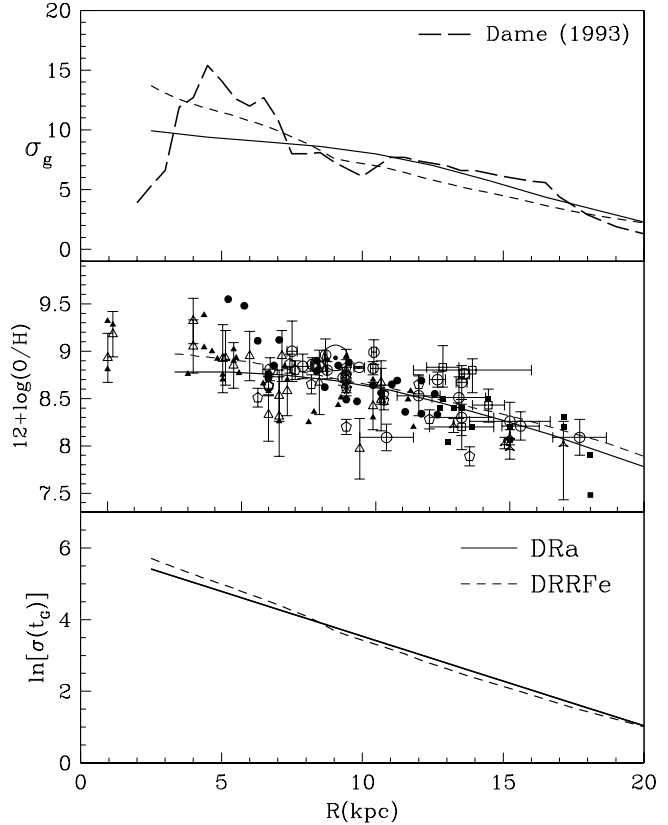,width=8.9truecm}}
\caption{A ``successful'' {\sf DRRF} model compared to the case with no flows.}
\label{drrfbestfit}
\end{figure}

To reproduce the observed metallicity gradient with this SF law,
negligible inflow is required in the outer regions where static models
already predict a gradient with the right slope (see model {\sf DRa}), 
while a moderate inflow velocity ($\sim$0.3~km~sec$^{-1}$) in the inner parts 
is needed to maintain the observed slope also where the predicted gradient 
would otherwise flatten (see PC99). Such a model is, for example, model
{\sf DRRFe}, obtained with the inflow velocity pattern shown in 
Fig.~\ref{velTA15bestfit}.
In Fig.~\ref{drrfbestfit} the ``successful'' model {\sf DRRFe} is compared 
to the original model {\sf DRa} of PC99 with no radial flows.

\subsection{Concluding remarks}

The aim of the previous ``successful'' examples is to remark how, for each 
SF law considered,
it is possible to tune the inflow velocity pattern so as to get
a good overall agreement with the observational
data. As an indication, suitable inflow profiles have the following 
characteristics.
\begin{itemize}
\item
With a Schmidt SF law, relatively fast (but still plausible)
flows are needed within the star--forming disc, with velocities of the order
of 1~km~sec$^{-1}$. The required metallicity gradient is easily obtained,
especially if the inflow velocity is slowly decreasing inward. \\
Inflow from the outer gaseous disc must be less prominent (though present), 
otherwise the predicted gas density in the outer regions is too high.
\item
In models adopting an Oort--type SF law with $\kappa=1.0$, a good fit to the
data is obtained with an inflow profile of the same kind:
inflow at $\sim 0.5$~km~sec$^{-1}$ from the outer gaseous disc,
drift velocities raising at $\sim 1$~km~sec$^{-1}$ at the stellar disc edge
and declining inward. Curiously enough, in this case
the higher SF efficiency in the inner regions combines with the slowdown of 
inflows to predict a peak in the gas distribution around $r=3$~kpc, quite
reminiscent of the observed molecular ring.
\item
Models adopting the SF law by Dopita \& Ryder give a good description of the 
outer regions of the Disc already in the absence of flows (PC99). Mild 
radial inflows can be assumed in the inner 
regions, where the radial gradient would otherwise flatten, to obtain the
right slope throughout the disc. The required drift velocities are around
0.3~km~sec$^{-1}$, which can be reasonably provided, for instance, by the 
shocks occurring within the spiral arms (see \S\ref{previous}).
\end{itemize}
The various ``successful models'' presented here are not meant to be
definitive recipes to reproduce the 
radial profile of the Galactic Disc. In fact, the adopted inflow velocity 
profiles are quite
arbitrary and {\it ad hoc}. These models are rather meant to show how radial 
flows can be a viable mechanism to interpret the properties of the 
Disc. In PC99 we showed how none of the various SF laws investigated
is able, by itself, to reproduce the observed metallicity 
gradient throughout the whole extent of the Disc, unless some additional 
``dynamical'' assumption is included. In PC99 we considered 
the classical case of an inside--out disc formation, namely of an
infall time-scale increasing outward. Here, we just show that radial inflows 
can provide another viable 
``dynamical'' assumption to be combined with any SF law to reproduce the 
observed gradient. In particular, if we adopt a Schmidt SF law 
with $\kappa=1.5$ or an Oort--type SF law with $\kappa=1.0$, as
recent empirical evidence seems to support (Kennicutt 1998), the
required variation of the accretion time-scale is too extreme in the
pure inside--out assumption (PC99), and radial inflows are then 
necessary to explain the metallicity gradient. 
Of course, the two effects (inside--out formation and 
radial flows) can also play at the same time. We do not address here
their combined outcome, because in our models this would merely translate into 
increasing the number of parameters which one can tune to fit the observational
data. No further insight in the problem would be gained. With this kind of
models we can just learn how the various players (different SF laws,
radially varying accretion time-scales, radial gas inflows) enter the game of
reconstructing the general picture, and analyse their effects one by one.

It is also worth stressing that even slow radial gas flows, with velocities 
well plausible in terms of the triggering physical mechanisms and
within the observational limits (\S\ref{previous}), 
have non--negligible influence on chemical
models, especially on the gas density distribution. It is therefore misleading
to seek for a one--to--one relation between gas content and metallicity, or 
between gas profile and metallicity gradient, like the one predicted by simple
models (e.g.\ Tinsley 1980). When studying the chemical features of galaxies, 
it is very dangerous to assume that such a relation must hold, since even mild
flows can easily alter the overall distribution.

The models with smooth radial inflows presented in this section (with the 
possible exception of model {\sf O10RFe}) are still unable to reproduce the 
gas density peak corresponding to the molecular ring around 4~kpc, since that
needs to take into account the peculiar dynamical influence of the Galactic Bar
on gas flows. This will be the issue of the next section.


\section{The role of the Galactic Bar}
\label{bar}

There is by now substantial evidence that the Milky Way hosts a small Bar 
in its inner 3~kpc or so. The idea was originally suggested to explain the
kinematics of the atomic and molecular gas near the Galactic Centre
(de Vaucouleurs 1964, Peters 1975, Liszt \& Burton 1980, Mulder \& Liem 1986).
In recent years further evidence for a Galactic Bar has piled up from several
tracers: gas dynamics (Binney \etal 1991; Weiner \& Sellwood 1996, 1999;
Yuan 1993; Wada \etal 1994; Englmaier \& Gerhard 1999; Fux 1999), IR photometry
and star counts (Blitz \& Spergel 1991, Weinberg 1992, Nikolaev \& Weinberg 
1997, Dwek \etal 1995, Binney \etal 1997, Unavane \& Gilmore 1998), stellar 
kinematics (Zhao \etal 1994; Weinberg 1994; Fux 1997; Sevenster 1997, 1999;
Sevenster \etal 1999; Raboud \etal 1998), OGLE data (Stanek \etal 1994, 1997; 
Paczynski \etal 1994; Evans 1994;
Zhao \etal 1995, 1996; Zhao \& De Zeeuw 1998, Ng \etal 1996). For a 
review on the Galactic Bar see e.g. Gerhard (1996, 1999). 
The determination of the characteristic parameters (size,  
axis ratio, rotation speed, orientation and so forth) is even more
difficult for the Bar of our own Milky Way than for external galaxies. Broadly
speaking, the various studies mentioned above indicate a Bar with a major axis
of {\mbox{2--4~kpc}} viewed at an angle of {\mbox{15--45$^o$}} in the first 
longitude quadrant, an axis ratio around 3:1 and a pattern speed 
{\mbox{$\Omega_p \sim 60$~km sec$^{-1}$ kpc$^{-1}$}}.

PC99 underlined that the dynamical influence of the Galactic Bar is likely to
account for the peak at 4~kpc displayed by the gas profile in the Disc, which
static chemical models are unable to reproduce if other constraints, like the
observed metallicity gradient, are to be matched as well. In fact, 
gravitational torques in a barred, or non--axisymmetric, potential are thought
to induce gas accumulation and formation of rings at the corresponding 
Lindblad resonances (e.g. Combes \& Gerin 1985; Schwarz 1981, 1984).
In brief, bar--induced flows sweep gas away from the co--rotation (CR) radius,
where the bar roughly ends, toward the inner and outer Lindblad resonances 
(ILR, OLR).
In fact, we developed our new chemical model with radial flows also with 
the aim to mimic such effects of the Bar upon the gas distribution, by
simulating suitable flow velocity profiles.

As mentioned above, though the existence and gross features of the 
Galactic Bar are by now established, there is no
general agreement on details like its size and pattern speed, and on the
corresponding radii for its CR and ILR, OLR. In this paper, with 
the aim to reproduce the
molecular ring at 4--6~kpc, we will consider two Bar models covering
the range of scenarios suggested in literature:
\begin{description}
\item[{\bf case A)}]
Bar's CR around 3.5~kpc, so that the dip in the gas distribution between 1.5
and 3.5~kpc is interpreted as a fast inward drift of gas from CR to the ILR
and the nuclear ring;
\item[{\bf case B)}]
Bar's CR around 2.5~kpc and OLR around 4.5~kpc, so that the molecular ring is
interpreted as accumulation of gas from CR toward the OLR.
\end{description}
In any case, we will assume here that the Bar 
influences
only the inner {\mbox{5--6~kpc} of the Galactic Disc, where its OLR is 
supposed to lie at the outermost according to current understanding, 
while leaving regions outside the OLR unaffected (see also Gerhard 1999).
Actually, the possible influence of the Bar over a larger Disc region
than its formal extent is still an open problem, as we will 
comment upon in the final conclusions (\S\ref{conclusions}).

In the framework of our chemical model, the inclusion of the Bar translates 
into imposing a suitable velocity profile for the radial flows 
in the inner regions of the disc. 
Namely, we run the ``successful models'' with radial flows 
presented in \S\ref{bestfit}, but at a
suitable age the Bar is assumed to form and the radial velocity profile
is altered correspondingly, by modifying
the coefficients $\alpha_k$, $\beta_k$, $\gamma_k$ describing the radial flow 
pattern (\S\ref{discrete}).
In case A, we will impose fast radial inflow velocities within CR at 3--4~kpc
to mimic the rapid drift of the gas toward the ILR (\S\ref{caseA}). 
In case B, we will impose outflows from CR to the OLR around 4.5~kpc
(\S\ref{caseB}).

\begin{figure}[t]
\centerline{\psfig{file=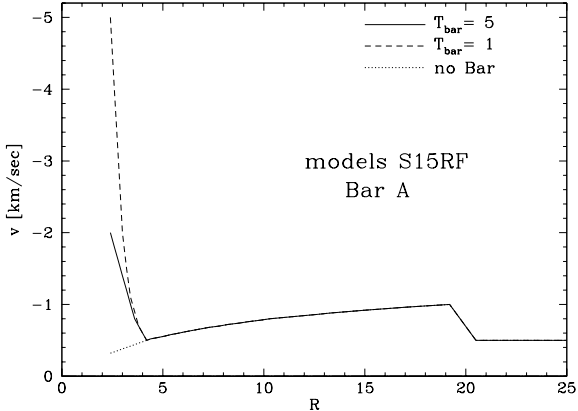,angle=-90,width=8truecm}}
\caption{Inflow velocity pattern adopted from the onset of the Bar at the time
$t_G-T_{bar}$ upon the base ``no Bar'' model ({\sf S15RFe}), to mimic case
A for the Galactic Bar with models {\sf S15RF}. The dotted line applies to the 
base model {\sf S15RFe}, and also to the ``barred'' models before
$t_G-T_{bar}$.}
\label{vels15barA}
\end{figure}
%
\begin{figure}[h]
\centerline{\psfig{file=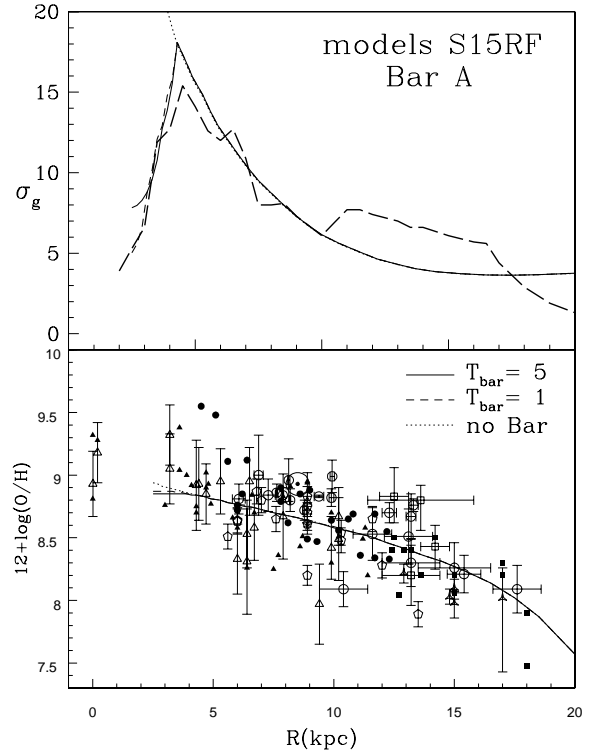,width=8truecm}}
\caption{Models {\sf S15RF} mimicking case A for the Galactic Bar, adopting
the inflow patterns of Fig.~\protect{\ref{vels15barA}}. The ``no Bar'' model 
{\sf S15RFe} is also shown for comparison as a dotted line.}
\label{s15barA}
\end{figure}

As to the age of the Bar, an upper limit is set by the typical age of its 
stellar
population, 8--9~Gyr (Ng \etal 1996); an age of 5 to 8~Gyr has been suggested 
by Sevenster (1997, 1999). The results from our simulations turned out to be
quite
insensitive to a change in the Bar's age from 9 to 5~Gyr; therefore, we will
present simulations with a 5~Gyr old Bar 
($T_{bar}=5$) 
as representative of the generic case of age $\gsim 5$~Gyr.
We will also consider, for the sake of completeness, the case of a much
younger Bar of 1~Gyr of age ($T_{bar}=1$).


\subsection{Modelling the effects of the Bar: case A}
\label{caseA}

To mimic case A, where the Bar induces fast inflows from CR to its ILR,
the inflow velocity is 
typically increased for $r < 3.5-4$~kpc (the Bar's CR radius) with respect 
to the drift pattern adopted before the onset of the Bar.
Models corresponding to case A give good results when combined with a Schmidt
SF law with radial inflows. Starting from the corresponding ``successful'' 
model {\sf S15RFe} of \S\ref{bestfit} and changing the velocity law as in 
Fig.~\ref{vels15barA} at the onset of the Bar, we obtained
the models shown in Fig.~\ref{s15barA}, compared to the original model 
{\sf S15RFe} with no Bar's effects included. Obviously, if the Bar is younger
($T_{bar}=1$~Gyr) faster induced inflows need to be assumed in the
inner regions so that the observed sharp dip in the gas profile at 
$r \lsim 3.5$~kpc is obtained in a shorter time (Fig.~\ref{vels15barA}).
Anyways, no extreme speeds need to be induced by the Bar ($|v| < 5$~km~sec) 
at these radii yet, to resemble the observed gas profile, so the models remain
plausible.

This type of solution 
actually corresponds to 
the one originally suggested by Lacey \& Fall (1985), who in fact assumed a
Schmidt SF law, radial inflows of the order of --1~km~sec$^{-1}$ to reproduce 
the metallicity gradient, and for $r \leq 4$~kpc a raise in the inflow speed 
up to {\mbox{--10~km~sec$^{-1}$}} to reproduce the gas profile.
Without such a spike in the inflow velocity, the gas distribution keeps rising 
inward, with no depression (cfr.\ model {\sf S15RFe}).
This picture also corresponds to the situation for viscous models as suggested
by Thon \& Meusinger (1998): the detailed gas profile of the Disc 
can be reproduced only by artificially increasing the viscosity
in the inner regions, so as to mimic the influence of the Galactic Bar. 

\begin{figure}[ht]
\centerline{\psfig{file=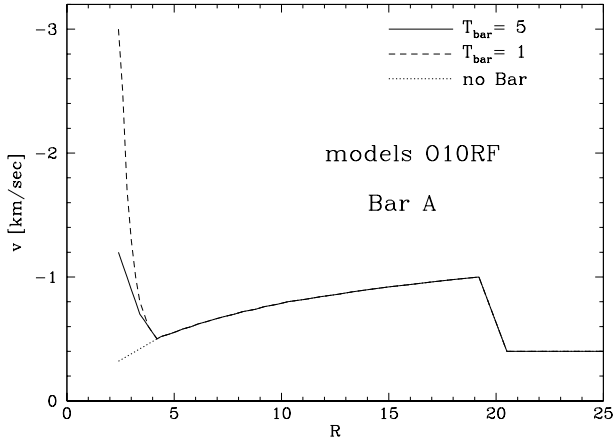,angle=-90,width=8.5truecm}}
\caption{Inflow velocity pattern adopted from $t_G-T_{bar}$
upon the base (``no Bar'') model {\sf O10RFe}, to mimic
case A for the Galactic Bar with models {\sf O10RF}.}
\label{velws10barA}
\end{figure}
%
\begin{figure}[ht]
\centerline{\psfig{file=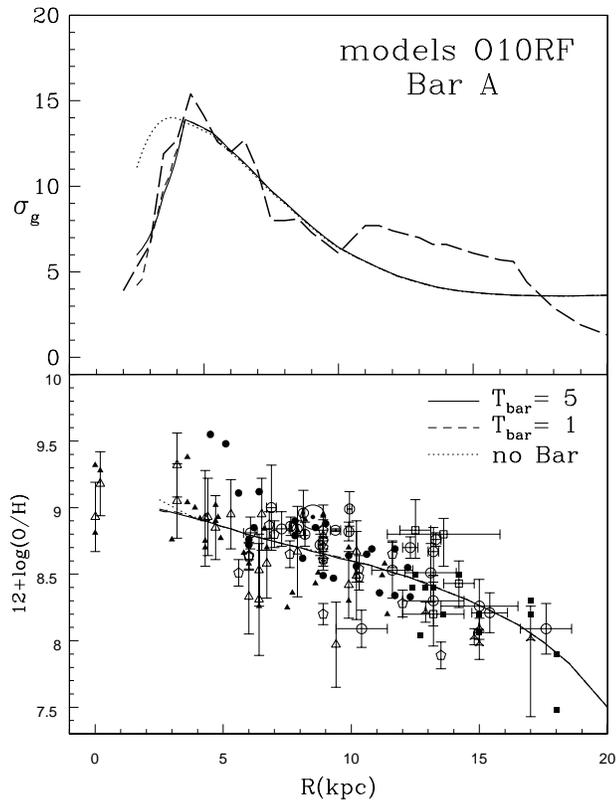,width=8.5truecm}}
\caption{Models {\sf O10RF} mimicking case A for the Galactic Bar, adopting
the inflow patterns of Fig.~\protect{\ref{velws10barA}}.}
\label{ws10barA}
\end{figure}

Case A can also well combine with model {\sf O10RFe}, namely with an 
Oort--type SF law with $\kappa=1$ plus the corresponding ``successful'' inflow
pattern. In fact, also in model {\sf O10RFe} (which is actually the only model
able to predict a peak in the gas profile reminding of the molecular ring 
even without including the Bar, see \S\ref{bestfit}) the gas profile 
rises inward,
and by increasing the inflow velocities inside 4~kpc one can fit the observed
gas depression. As an example, we show models {\sf O10RF} with inflow patterns
as in  Fig.~\ref{velws10barA}; the corresponding results are plotted in 
Fig.~\ref{ws10barA} together with the base model {\sf O10RFe}.

Notice how in all these ``case A'' models the metallicity gradient is only 
negligibly affected by the switch of the inner inflow pattern 
at the time of onset of the Bar, $t_G-T_{bar}$ ($t_G=15$~Gyr is the present 
Galactic age). The effects are at most limited to $r \lsim 3$~kpc, where
they are hard to check from observations, since in that region
the gas distribution is depressed and 
the metallicity tracers are missing as well, since they are young objects 
strongly correlated to present--day SF activity and therefore 
to the presence of gas. We remark that abundance data at $r=0$
in the plots for the abundance gradient are to be disregarded 
as a constraint for the model, since
they refer to the Galactic Centre population, not to the Disc (see PC99).


\subsection{Modelling the effects of the Bar: case B}
\label{caseB}

According to what we labelled above as ``case B'', the Galactic Bar ends 
in correspondence to its CR around 2.5~kpc, and it has an OLR around 4.5~kpc. 
Therefore, the gas is expected to drift from CR outward
and accumulate at the OLR, while gas inflowing from outer regions slows down
and accumulates as well at the level of the resonance at 4.5~kpc.

\begin{figure}[b]
\centerline{\psfig{file=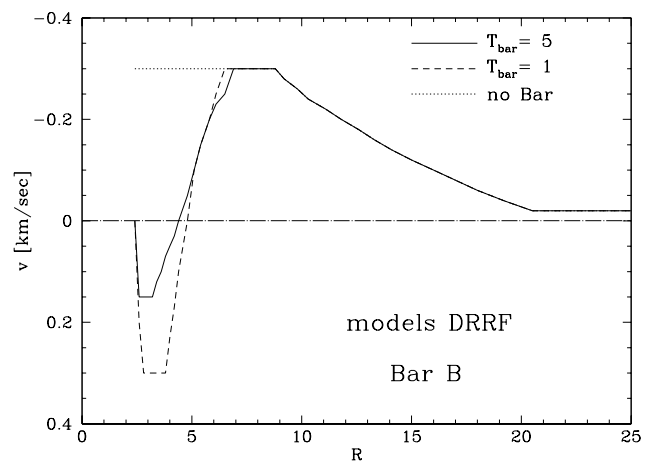,angle=-90,width=8.7truecm}}
\caption{Radial flow velocity pattern adopted from $t_G-T_{bar}$
upon the base ``no Bar'' inflow pattern, to mimic case B 
for the Galactic Bar with models {\sf DRRF}. The dot--dashed 
horizontal line at $v=0$ marks the transition from inflows (negative $v$) to 
outflows (positive $v$).}
\label{velta15barB}
\end{figure}

Models for case B will adopt, starting from $t_G-T_{bar}$, a radial
flow pattern with positive velocities (outflow) from $r=2.5$~kpc to
$r \sim 4.5$~kpc, and negative velocities (inflow) from outside dropping to 
zero at $r \sim 4.5$~kpc.
Case B can be combined with model {\sf DRRFe}, namely with the SF law by 
Dopita \& Ryder (1994) and the corresponding ``successful'' overall inflow 
pattern (\S\ref{bestfit}). The relevant models with their detailed velocity 
patterns, for the usual two values of $T_{bar}$, are shown in 
Figs.~\ref{velta15barB} and~\ref{drbarB}. 
Notice once again that rather low drift velocities 
{\mbox{($|v| < 0.5$~km~sec)}}
suffice to reproduce the gas peak, which reinforces the plausibility of the
models. 

In case B models, the metallicity gradient is clearly affected in the
inner regions by the switch in the gas flow pattern, at least if this lasted
for some time (Fig.~\ref{drbarB}, case $T_{bar} = 5$). This feature,
though, is again hard to check from observations since there are no tracers of
Disc metallicity for $r < 4$~kpc (see \S\ref{caseA}).

\begin{figure}[t]
\centerline{\psfig{file=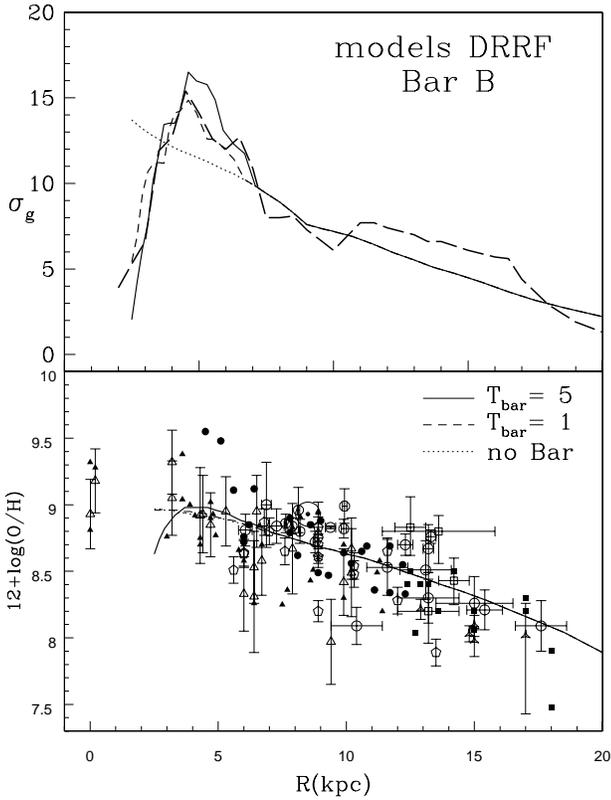,width=8.5truecm}}
\caption{Models {\sf DRRF} mimicking case B for the Galactic Bar, adopting
the flow patterns of Fig.~\protect{\ref{velta15barB}}. Disregard abundance 
data at $r=0$ (see text).}
\label{drbarB}
\end{figure}

Case B can also reproduce the observed gas profile combined with an Oort--type
SF law with $\kappa=1$ (model {\sf O10RFe}), especially provided
the Bar formed recently, as displayed in Fig.~\ref{ws10barB} with 
$T_{bar}=1$. If the Bar--induced flows activate much earlier, the predicted
peak is very sharp and narrow (Fig.~\ref{ws10barB}, case $T_{bar}=5$~Gyr),
due to the milder sensitivity of SF to the gas density in this
case (Schmidt-like exponent {\mbox{$\kappa=1$}}): with respect to other 
SF laws, this SF is relatively less efficient where the gas density is high, 
namely where the gas accumulates around 4.5~kpc, while it is 
relatively more efficient where the density drops, below 4~kpc.
Such a sharp peak seems in contrast with the observed, quite
broad distribution (the observational uncertainty on the gas density
profile in the inner Galactic region is less than a factor of 2, Dame 1993).
Models {\sf O10RF} with an ``old'' Bar are therefore less appealing in case B.

\begin{figure}[t]
\centerline{\psfig{file=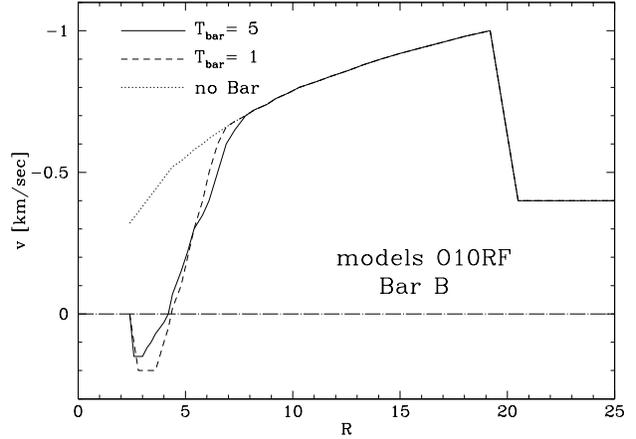,angle=-90,width=8.5truecm}}
\caption{Radial flow velocity pattern adopted from 
$t_G-T_{bar}$ upon the base ``no Bar'' inflow pattern, to mimic case B 
for the Galactic Bar with models {\sf O10RF}.}
\label{velws10barB}
\end{figure}
%
\begin{figure}[t]
\centerline{\psfig{file=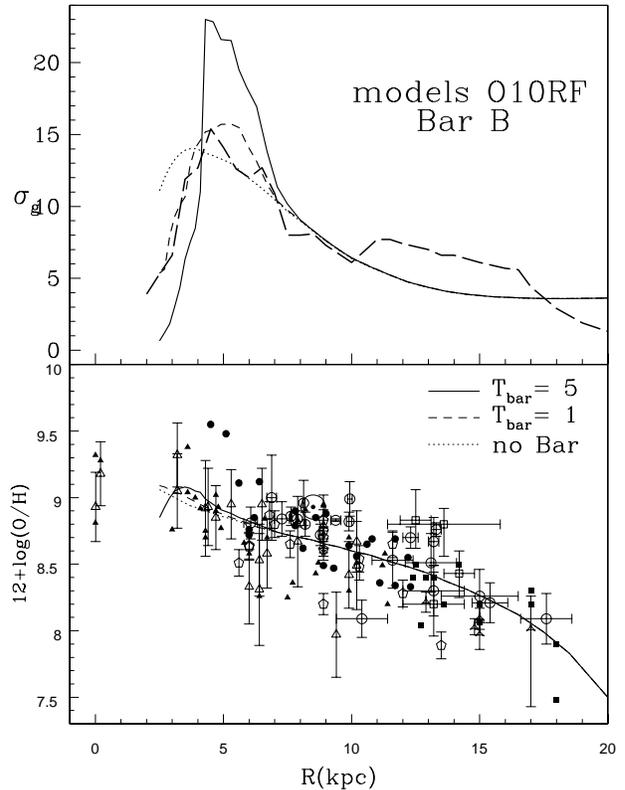,width=8.5truecm}}
\caption{Models {\sf O10RF} mimicking case B for the Galactic Bar, adopting
the flow patterns of Fig.~\protect{\ref{velws10barB}}.}
\label{ws10barB}
\end{figure}


\subsection{Concluding remarks}

We referred to the current understanding of the structure and
features of the Galactic Bar to simulate its dynamical influence on the 
surrounding gas flows and distribution. 
Chemical models accounting for the effect of the Bar make it finally possible
to reproduce the gas profile properly, which could not be accomplished by
static models (PC99).

Broadly speaking, two main scenarios are presented.
\begin{enumerate}
\item
The Bar stretches to 3.5~kpc and the gas peak is due to a rapid
depletion of gas drifting from the Bar's CR to the Galactic Centre (case A).
To reproduce the gas profile, we need a gas distribution which keeps
increasing from the outer disc inward, down to $\sim 4$~kpc where the Bar's
influence produces the sharp depletion. This can be achieved only by means of
efficient radial inflows over the whole disc, and in this case radial inflows
are also the major source for the observed metallicity gradients (models
{\sf S15RFe} and {\sf O10RFe}).
\item
The Bar ends around 2.5~kpc, where its CR is set, and the gas peak is due 
to the accumulation of gas from CR outward to its OLR at $\sim 4.5$~kpc 
(case B). A rather 
limited contribution of radial inflows from the outer regions suffices to 
obtain the peak. In this case, the metallicity gradients in the outer regions 
of the disc may be mainly due to the SF process itself, and to the intrinsic 
variation of the SF efficiency with radius (model {\sf DRRFe}).
\end{enumerate}

Within these simplified models it is unfeasible to discuss any further
on the scenarios for Bar structure and age, and 
related gas flows. Only detailed dynamical simulations for Bar formation, 
evolution and potential can tell how the molecular ring consequently formed,
which is beyond the goals of this paper (see also \S\ref{conclusions}). 
Here we were just interested in showing how even simple qualitative models for 
Bar--induced gas flows in the inner disc solve in fact the puzzle encountered
in PC99. Namely, they can reproduce at the same time both 
the metallicity gradient and the gas distribution, in particular the peak 
corresponding to the molecular ring around 4~kpc. This can be accomplished
already with quite slow, and largely plausible, flow velocities.
The present modelling therefore provides a simple tool for qualitative 
understanding of possible behaviours.

Regardless of details, however, one condition is necessary for the scheme 
to work: there must be enough gas in the inner regions of the Disc, which the 
Bar can then ``shape'' to resemble the detailed observed distribution. This 
favours chemical models with radial inflows where the metallicity
gradient can coexist with high gas fractions in the inner shells. Our
simulations showed that no Bar--induced gas flow
superimposed on otherwise static models (as those by PC99) can produce 
the observed gas peak: whatever the 
assumed age of the Bar or gas velocity profile, there is not enough gas
left in the inner Galactic regions if we are to 
reproduce the metallicity gradient as well. Gas must be continuously 
replenished by inflows from outer regions;
that's why we presented here
``barred'' models based only on the ``successful'' models with radial inflows 
from \S\ref{bestfit} and never on the static models of PC99.


\section{Summary and conclusions}
\label{conclusions}

From the results of static chemical models, PC99 underlined the need to 
introduce radial flows to explain some features of the Galactic Disc. In fact,
static models are unable to reproduce, at the same time, both the
metallicity gradient and the radial gas profile; in particular, the peak
corresponding to the molecular ring at 4--6~kpc is likely to be a consequence 
of gas drifts induced by the dynamical influence of the Galactic Bar.
Therefore, in the present paper we introduced a new chemical model including 
radial gas flows, developed as a multi--dimensional generalization of the 
original static model (\S\ref{discrete}).
Our model is conceived so as to adapt to any imposed radial velocity 
profile, describing both inflows and outflows in any part of the disc. 
The model is carefully tested 
against instability problems and 
spatial resolution, by comparing it to suitable exact analytical cases
(Appendix~B). In this paper we applied the model to the Galactic Disc;
more in general, such models allowing for gas drifts are meant to be used
as fast and handy interfaces between detailed dynamical galaxy
models (predicting the velocity profiles) and parametric chemical
and spectro--photometric models.

An overview of the behaviour of chemical models with radial inflows of gas
shows that these provide an alternative ``dynamical'' assumption to the
inside--out disc formation scenario to explain the metallicity gradient
(\S\ref{radfloweffects}).
With radial gas flows, the model can reproduce the metallicity
gradient even in the case of a Schmidt or an {\mbox{Oort--type}} SF law,
which were excluded in the case of static models (see PC99).
In addition, it appears that even low radial flow velocities,
well within observational limits and theoretical expectations 
(see~\S\ref{previous}), have non--negligible 
effects upon model predictions on the metallicity
gradient and moreover on the gas distribution. In particular,
if radial gas inflows are allowed for, a metallicity gradient can coexist 
with a high gas fraction in the inner regions, at odds with simple static 
models. This is indispensable to reproduce the observed gas distribution 
in the inner Galaxy (see point 2 below).
The remarkable effects of even slow radial flows upon observable quantities,
mainly upon the gas distribution, should be kept in mind as a caveat when 
comparing real galaxies to simple analytical models which predict a 
one--to--one relationship between metallicity and gas fraction (e.g.\ Tinsley 
1980). Our models show that small dynamical effects, like slow gas flows, 
can easily make real systems depart from the behaviour of simple models.

With our model it is possible to mimic the dynamical influence of the
Galactic Bar and reproduce the peak in the gas distribution around 4~kpc
(\S\ref{bar}).
Two scenarios, related to two different models for the structure of the
Bar, are qualitatively suggested.
\begin{description}
\item[{\bf A.}]
With a Schmidt or an Oort--type SF law, slow radial inflows in the disc 
pile up gas inward down to {\mbox{$r=3.5-4$~kpc}}. Here, the Bar CR
radius is found and the gas is quickly swept inward from CR toward an ILR, 
which causes the drop in the gas profile at 3.5~kpc.
\item[{\bf B.}]
With a SF law like that by Dopita \& Ryder (1994),
smaller inflow rates suffice
to reproduce the metallicity gradient, leading to a lower concentration of gas 
in the inner regions than in the previous case. The peak corresponding to the
molecular ring can be reproduced with a Bar CR around 2.5~kpc and its
OLR around 4.5~kpc, so that all the gas external to $r=2.5$~kpc tends to pile
up around the OLR.
\end{description}
Though these models are just qualitative and cannot
describe the detailed dynamical process of Bar formation nor the
evolution of the related gas flows to form the molecular ring, they provide
two interesting indications. 
\begin{enumerate}
\item
Only when introducing the effects of the Bar, the model is able to reproduce 
the radial gas profile properly. The only possible exception resides in a 
particular combination of an Oort-type SF law with a radial inflow pattern 
whose velocity decreases inward (model {\sf O10RFe}; this combination may lead
to a peak of the gas distribution in the inner Galactic regions,
closed to the observed molecular ring. But this particular, fortunate case 
does not diminish the general conclusions about the role of the Galactic Bar.
\item
In any case (A or B above), overall radial inflows in the disc are 
indispensable to replenish the inner regions with enough gas that the observed
molecular ring can form
under the influence of the Bar. This seems to favour disc models with radial 
inflows, unless one assumes that the gas in the ring has 
some different origin (gas swept from the Bulge, or accreted later).	
\end{enumerate}
To investigate these issues any further, detailed gas--dynamical models
are obviously required. Unfortunately, most studies on Bar--induced 
gas dynamics 
(see references in \S\ref{bar}) concentrate on the observed features of the
very inner regions, such as the nuclear ring, the 3~kpc expanding arm, and so
forth. Little discussion can be found about the effects of the Bar on more
external regions, and on the formation of the molecular ring in particular:
whether it is due to gas depletion inside CR as in our case A, or due to
gas accumulation at some resonance (e.g.\ Binney \etal 1991, Fux 1999) as in 
our case B, or whether it just consists of two or more tightly wound spiral
arms (e.g. Englmaier \& Gerhard 1999). Further gas--dynamical studies  
suggesting detailed scenarios, time-scales, and velocity profiles
for the formation of the molecular ring would be welcome, for the sake of 
including the effects of the Bar in chemical evolution 
models more consistently. 

Further investigation of gas--dynamical models on the influence 
of the Bar on even larger scales (namely, outside its OLR) should be pursued
as well, since this is a clue issue related to
a claimed discrepancy between
the characteristics of the Galactic Bar and the observed metallicity gradient.
It is well known that barred galaxies display systematically shallower
gradients than ordinary spirals (e.g.\ Alloin \etal 1981, Vila--Costas \& 
Edmunds 1992, Martin \& Roy 1994). This is likely a consequence of
the radial mixing induced by bars; in fact, Martin \& Roy (1994) found
a correlation for external galaxies between the strength of a bar 
and the metallicity gradient. Taking this empirical relation at face value, 
the Galactic Bar with an axial ratio of {\mbox{$\sim 0.5$}} should induce 
a metallicity gradient of {\mbox{--0.03~dex/kpc}}, much shallower than the 
observed one of --0.07~dex/kpc, which is typical of a {\it normal} Sbc galaxy.
To overcome such a puzzle, it has been suggested that the Galactic Bar must
be very young ($<$1~Gyr), so that there was not enough time yet to flatten 
the gradient (Gummersbach \etal 1998); but this is in conflict with other
estimates of the Bar's age (e.g.\ Sevenster 1997, 1999). Alternatively,
we suggest that the discrepancy might be only apparent, since the Galactic Bar
is quite small, and the Milky Way cannot be properly considered a barred 
spiral.
It might be unlikely that the Bar can influence the metallicity 
gradient all over the Disc, as in really barred galaxies: Bar--induced 
radial drifts and corresponding chemical mixing are expected to occur from CR 
toward the ILR (inflows) and to the OLR (outflows; e.g.\ Schwarz 1981, 1984; 
Friedli \etal 1994).
Present understanding of the Galactic Bar sets its OLR between 4.5 and 6~kpc
(\S\ref{bar} and references therein), so in our models we presumed 
that the Bar induces negligible mixing beyond these radii, regardless
of its age (see also Gerhard 1999). If the situation is as in the models 
we presented here, in fact,
the metallicity gradient in the outer regions
is unperturbed and just related to intrinsic Disc properties and/or 
large--scale viscous flows.

However, gas--dynamical simulations dedicated to the effects of the Galactic
Bar over the whole Disc would be necessary, so as to investigate 
the relation between the Bar, radial mixing and the metallicity 
gradient more consistently.
More in general, including the effects of {\mbox{bar--induced}} 
radial flows in the 
picture of the chemical evolution of spiral galaxies might turn out to be
of wide interest, since it is likely that all spirals develop at some point, 
or have developed in the past, some bar--like structure (Binney 1995). 
Infrared observations indeed reveal that a large fraction of spirals 
host a barred structure (e.g.\ Eskridge \etal 1999), and 
recent numerical simulations suggest that even weak bars or oval distortions 
may be able to induce radial drifts to form multiple gaseous rings at the 
corresponding Lindblad resonances (Jungwiert \& Palou\v{s} 1996). 
Bars could even drive secular evolution of spiral discs from late to early type
(e.g.\ Dutil \& Roy 1999). 
Bar--driven radial gas flows might therefore play a fundamental role 
in the chemical evolution of spiral discs.


\acknowledgements{We thank Joachim K\"oppen for his advice on 
numerical modelling of radial gas flows, Yuen K.\ Ng and Antonella Vallenari
for useful discussions
about Galactic structure, and our referee, Mike Edmunds, whose suggestions 
much improved the presentation of our paper. 

\noindent
L.P.\ acknowledges kind hospitality from the 
Nordita Institute in Copenhagen, from the Observatory of Helsinki and from 
Sissa/Isas in Trieste. This study has been financed by the Italian 
MURST through a PhD 
grant and the contract ``Formazione ed evoluzione delle galassie'', 
n.~9802192401.}


\appendix
\section{The explicit expression of Eq.~(\protect{\ref{solsystemradf2}})}
\label{appendixA}

Here we sort out the explicit expression~(\protect{\ref{solradfN}})
of the solution~(\protect{\ref{solsystemradf2}}) for $G_i(r_k,t)$, by 
calculating the
matrix $e^{t {\cal A}}$. Let us first introduce the simplified notation:
\begin{equation}
\label{notation}
G_k(t) = G_i(r_k,t),~~~~~~~~~~~\overline{W_k} = \overline{W}_i(r_k)
\end{equation}
For the sake of example, we present here the detailed solution in the case 
of three shells ($N=3$). The relevant system of 
equations~(\protect{\ref{systemradf}}) becomes:
\begin{equation}
\label{systemradf3}
\left\{ \begin{array}{l}
\frac{d}{dt} \, G_1(t) = \vartheta_1 G_1(t) + \gamma_1 G_2(t) + W_1 \\
\\
\frac{d}{dt} \, G_2(t) = \alpha_2 \, G_1(t) + \vartheta_2 G_2(t) +
			 \gamma_2 G_3(t) + W_2 \\
\\
\frac{d}{dt} \, G_3(t) = \alpha_3 \, G_2(t) + \vartheta_3 G_3(t) + W_3 +
\overline{\omega_i}\\
\end{array} \right.
\end{equation}
where all the coefficients are considered as constants 
(see \S\ref{numericalradf}), though we have 
omitted the bar over $W_k$ and $\vartheta_k$ for simplicity. 
A system like~(\protect{\ref{systemradf3}}) holds for each chemical species 
$i$, but the characteristic matrix ${\cal A}$ is the same for any $i$
(see \S\protect{\ref{numericalradf}}); in this case:
\[ {\cal A} = \pmatrix{
\vartheta_1 &   \gamma_1  &       0     \cr
  \alpha_2  & \vartheta_2 &   \gamma_2  \cr
      0     &   \alpha_3  & \vartheta_3 \cr} \]
Let's first notice, from the definition~(\protect{\ref{coeffradf}}),
that $\gamma_k$ and $\alpha_{k+1}$ can never be both positive: at least one of 
them must be zero since
they are ``activated'' in the opposite cases of inflow or outflow at
$r_{k+\frac{1}{2}}$, respectively; if there is no flow at all through 
$r_{k+\frac{1}{2}}$, they both reduce to zero. Therefore, in our calculations
we are always entitled to use the condition:
\begin{equation}
\label{gammacondition}
\gamma_k \, \alpha_{k+1} = 0 ~~~~~~~~~~~~~~~~~~\forall k
\end{equation}
The eigenvalues of the matrix ${\cal A}$ are
\[ \lambda_{1,2,3} = \vartheta_{1,2,3} \]
and the associated eigenvectors are:
\[ {\bf u}_1 = \pmatrix{
1 \cr
\cr
\frac{\alpha_2}{\vartheta_1 - \vartheta_2} \cr 
\cr
\frac{\alpha_3}{\vartheta_1 - \vartheta_3} 
\frac{\alpha_2}{\vartheta_1 - \vartheta_2} \cr} 
{\bf u}_2 = \pmatrix{
\frac{\gamma_1}{\vartheta_2 - \vartheta_1} \cr
\cr
1 \cr
\cr
\frac{\alpha_3}{\vartheta_2 - \vartheta_3} \cr} 
{\bf u}_3 = \pmatrix{
\frac{\gamma_1}{\vartheta_3 - \vartheta_1}
\frac{\gamma_2}{\vartheta_3 - \vartheta_2} \cr
\cr
\frac{\gamma_2}{\vartheta_3 - \vartheta_2} \cr
\cr
1 \cr} \]
From~(\protect{\ref{e^tAformula}}) we get:
\[ 
e^{t {\cal A}} 
\\
= {\scriptsize \pmatrix{ 
e^{\vartheta_1 t} & \gamma_1 f_{21}(t) &
\frac{\gamma_1 \gamma_2}{\vartheta_3-\vartheta_2} [f_{31}(t)-f_{21}(t)] \cr
 & \cr
\alpha_2 f_{21}(t) & e^{\vartheta_2 t} & \gamma_2 f_{32}(t) \cr
 & \cr
\frac{\alpha_2 \alpha_3}{\vartheta_2-\vartheta_1} [f_{32}(t)-f_{31}(t)] &
\alpha_3 f_{32}(t) & e^{\vartheta_3 t} \cr}}
\]
where we have indicated with:
\[ f_{kl}(t) = f_{lk}(t) \equiv
\frac{e^{\vartheta_k t}-e^{\vartheta_l t}}{\vartheta_k-\vartheta_l},
~~~~~~~~~~~~~~~~~~k,l=1,2,3 \]
Defining:
\[ g_k(\Delta t) \equiv \int_{t_0}^{t_1} e^{\vartheta_k (t_1-t)} \, dt \,=\, 
\frac{e^{\vartheta_k \Delta t} -1}{\vartheta_k} \]
the solution~(\protect{\ref{solsystemradf2}}) in the case $N=3$ becomes:
\begin{equation}
\label{solradf3}
\left\{ 
\begin{array}{l l}
G_1(t_1) = & e^{\vartheta_1 \Delta t} \, G_1(t_0) \,+ \\ 
 &     +\, \gamma_1 \, \frac{e^{\vartheta_2 \Delta t}-e^{\vartheta_1 \Delta t}}
            {\vartheta_2-\vartheta_1} \, G_2(t_0) \,+ \\ 
 &	+\, \gamma_1 \, \gamma_2 \, \frac{f_{31}(\Delta t)-f_{21}(\Delta t)}
	{\vartheta_3-\vartheta_2} \,G_3(t_0) \,+ \\
 &	+\,  W_1 \, \frac{e^{\vartheta_1 \Delta t} -1}{\vartheta_1} \,+ \\
 & 	+\, \gamma_1 \, \frac{W_2}{\vartheta_2-\vartheta_1} \left( 
            \frac{e^{\vartheta_2 \Delta t} -1}{\vartheta_2} -
            \frac{e^{\vartheta_1 \Delta t} -1}{\vartheta_1} \right) \,+ \\
 & +\, 	 \gamma_1 \gamma_2 \, \frac{W_3}{\vartheta_3-\vartheta_2} \left[
 \frac{g_3(\Delta t) - g_1(\Delta t)}{\vartheta_3-\vartheta_1} -
\frac{g_2(\Delta t) - g_1(\Delta t)}{\vartheta_2-\vartheta_1} \right] \\
 & \\
G_2(t_1)= & \alpha_2\, \frac{e^{\vartheta_2 \Delta t}-e^{\vartheta_1 \Delta t}}
           {\vartheta_2-\vartheta_1} \, G_1(t_0) \,+ \\
 & 	+\, e^{\vartheta_2 \Delta t} \, G_2(t_0) \,+ \\ 
 & +\, \gamma_2 \, \frac{e^{\vartheta_3 \Delta t}-e^{\vartheta_2 \Delta t}}
            {\vartheta_3-\vartheta_2} \, G_3(t_0) \,+ \\
 & +\, \alpha_2 \, \frac{W_1}{\vartheta_2-\vartheta_1} \left( 
	       \frac{e^{\vartheta_2 \Delta t} -1}{\vartheta_2} -
       \frac{e^{\vartheta_1 \Delta t} -1}{\vartheta_1} \right) \,+ \\ 
 & + \, W_2 \, \frac{e^{\vartheta_2 \Delta t} -1}{\vartheta_2} \,+ \\
 & + \,	 \gamma_2 \, \frac{W_3}{\vartheta_3-\vartheta_2} \left( 
\frac{e^{\vartheta_3 \Delta t} -1}{\vartheta_3} -
\frac{e^{\vartheta_2 \Delta t} -1}{\vartheta_2} \right)\\
 & \\
G_3(t_1) = & \alpha_2 \, \alpha_3 
\frac{f_{32}(\Delta t)-f_{31}(\Delta t)}{(\vartheta_2-\vartheta_1)}\,G_1(t_0)
      \,+ \\ 
 &   +\, \alpha_3 \, \frac{e^{\vartheta_3 \Delta t}-e^{\vartheta_2 \Delta t}}
	 {\vartheta_3-\vartheta_2} \, G_2(t_0) \,+ \\
 & +\,	e^{\vartheta_3 \Delta t} \, G_3(t_0) \,+ \\ 
 & 	+ \alpha_2 \alpha_3  \frac{W_1}{\vartheta_2-\vartheta_1} \left[
 	\frac{g_3(\Delta t) - g_2(\Delta t)}{\vartheta_3-\vartheta_2} -
   \frac{g_3(\Delta t) - g_1(\Delta t)}{\vartheta_3-\vartheta_1} \right] + \\
 & +\, \alpha_3 \, \frac{W_2}{\vartheta_3-\vartheta_2} \left( 
              \frac{e^{\vartheta_3 \Delta t} -1}{\vartheta_3} -
              \frac{e^{\vartheta_2 \Delta t} -1}{\vartheta_2} \right) \,+ \\
 & + \,	(W_3 + \overline{\omega_i}) \, 
	\frac{e^{\vartheta_3 \Delta t} -1}{\vartheta_3}
\end{array} 
\right.
\end{equation}
With zero flow velocity {\mbox{($\alpha_k=\beta_k=\gamma_k=0$),}}
(\protect{\ref{solradf3}}) 
reduces to the solving formula of the original static model (see PCB98).
Notice that the solution $G_1(t_1)$ for the $1^{st}$ shell includes not
only the contribution of the contiguous $2^{nd}$ shell, but also a contribution
from the $3^{rd}$ shell ``scaled'' by its passage through the $2^{nd}$ shell.
Similarly, the $3^{rd}$ shell is affected not only by the $2^{nd}$, but also
by the $1^{st}$ shell though they are not contiguous.

With analogous procedure, for an arbitrary number $N$ of shells the solution 
is of the kind:
\begin{equation}
\label{generalNsol}
G_k(t) = \sum_{l=1}^N F_{kl}(\Delta t) \, G_l(t_0) \,+\, 
   \sum_{m=0}^N H_{km}(\Delta t) \, W_m
\end{equation}
where:
\[ \begin{array}{l l r}
F_{kk}(\Delta t) = & e^{\vartheta_k \Delta t}\\
F_{kl}(\Delta t) = & \gamma_k \, \gamma_{k+1} ....... \gamma_{l-1} \,
{\cal F}_{k (k+1)....l}(\Delta t) & ~~~~l > k \\
F_{kl}(\Delta t) = & \alpha_{l+1} ....... \alpha_{k-1} \, \alpha_k 
{\cal F}_{l (l+1)....k}(\Delta t) & ~~~~l < k \\
H_{kk}(\Delta t) = & g_k(\Delta t) \\
H_{kl}(\Delta t) = & \gamma_k \, \gamma_{k+1} ....... \gamma_{l-1}
{\cal H}_{k (k+1)....l}(\Delta t) & ~~~~l > k \\
H_{kl}(\Delta t) = & \alpha_{l+1} ....... \alpha_{k-1} \, \alpha_k 
{\cal H}_{k (k-1)....l}(\Delta t) & ~~~~l < k
\end{array} \]
and the quantities ${\cal F}$ and ${\cal H}$ are constructed by means of 
recursive formul\ae:
\[ \begin{array}{l l}
{\cal F}_{ki}(\Delta t) = f_{ki}(\Delta t)~~~~~~~~~~~~~~~~~~~~ &
{\cal H}_{ki}(\Delta t) = 
\frac{g_k{\Delta t}-g_i{\Delta t}}{\vartheta_k-\vartheta_i} \\
 & \\
{\cal F}_{kij} = \frac{{\cal F}_{ki}-{\cal F}_{kj}}{\vartheta_i-\vartheta_j} &
{\cal H}_{kij} = \frac{{\cal H}_{ki}-{\cal H}_{kj}}{\vartheta_i-\vartheta_j}\\
 & \\
{\cal F}_{kijm} = 
\frac{{\cal F}_{kij}-{\cal F}_{kim}}{\vartheta_j-\vartheta_m} &
{\cal H}_{kijm} = 
\frac{{\cal H}_{kij}-{\cal H}_{kim}}{\vartheta_j-\vartheta_m}\\
 & \\
{\cal F}_{kijmn} = 
\frac{{\cal F}_{kijm}-{\cal F}_{kijn}}{\vartheta_m-\vartheta_n} &
{\cal H}_{kijmn} = 
\frac{{\cal H}_{kijm}-{\cal H}_{kijn}}{\vartheta_m-\vartheta_n}\\
 & \\
\vdots & \vdots
\end{array} \]
The coefficients $F_{kl}$ and $H_{kl}$ 
describe the contribution of the generic shell $l$ to the chemical
evolution of $k$. Notice that a shell external to $k$ ($l>k$) can influence $k$
only if all the inflow coefficients $\gamma$ in between $l$ and $k$ are 
non-zero, namely if there is a continuous inflow from $l$ to $k$, as expected.
The same holds for inner shells $l<k$, whose contribution $F_{kl}$, $H_{kl}$ 
is non-zero only if none of the intermediate outflow coefficients $\alpha$ is 
zero. 
The solution~(\protect{\ref{generalNsol}}) gets more and more 
complicated the larger the number $N$ of shells considered, since each shell 
formally feels the contribution of all the other shells
(as already noticed in the above case $N=3$). 
This occurs because~(\protect{\ref{generalNsol}}) 
would be the \underline{exact} analytical solution in the case of a
differential system with constant coefficients, namely \underline{if} the 
${\cal A}$ matrix in~(\protect{\ref{systemradf}}) were constant. Then, 
(\protect{\ref{generalNsol}})
would describe the complete evolution of any shell $k$, which over a 
galaxy's lifetime would indeed process and exchange gas drifting from or 
to rather distant shells. But ${\cal A}(t)$ is not constant even when 
the flow pattern $\alpha$, $\beta$, $\gamma$ is constant, because 
$\eta_k(t)$ and therefore $\vartheta_k(t)$ evolve in time due to SF; 
in fact, we apply 
(\protect{\ref{generalNsol}}) only upon short timesteps $\Delta t$, within 
which ${\cal A}(t)$ can be considered approximately constant.
If $\Delta t$ is short enough with respect to the radial flow velocities
--- as guaranteed by the Courant condition {\mbox{$\Delta t < v \Delta r$}},
see \S\protect{\ref{numericalradf}} ---,
we can assume that within $\Delta t$ the $k$-th shell is affected just by the
flows from the contiguous shells $k$+1 and $k$--1, and not from more distant
shells, although all of them formally contribute to the solution.
In this approximation we neglect all higher order terms in
$\alpha$ and $\gamma$, namely all the terms 
${\cal O}(\alpha_i \alpha_j)$ and ${\cal O}(\gamma_i \gamma_j)$, keeping
only the ``linear'' terms of the kind $\alpha_k f_{k (k-1)}$ and 
$\gamma_k f_{k (k+1)}$; so the general 
solution~(\protect{\ref{generalNsol}}) reduces in fact to~(\ref{solradfN}).


\section{Testing the numerical model}
\label{tests}

Since discretized numerical solutions for partial differential equations 
containing an ``advection term''
tend to be affected by instability problems (e.g.\ Press \etal 1986),
we tested our numerical code with gas flows against suitable analytical cases.
We report here two representative tests involving pure gas
flows (no SF) which allow for exact analytical solutions; to these
we compare the predictions of our numerical code with a SF efficiency 
dropped virtually to zero.

\begin{figure}[ht]
\centerline{\psfig{file=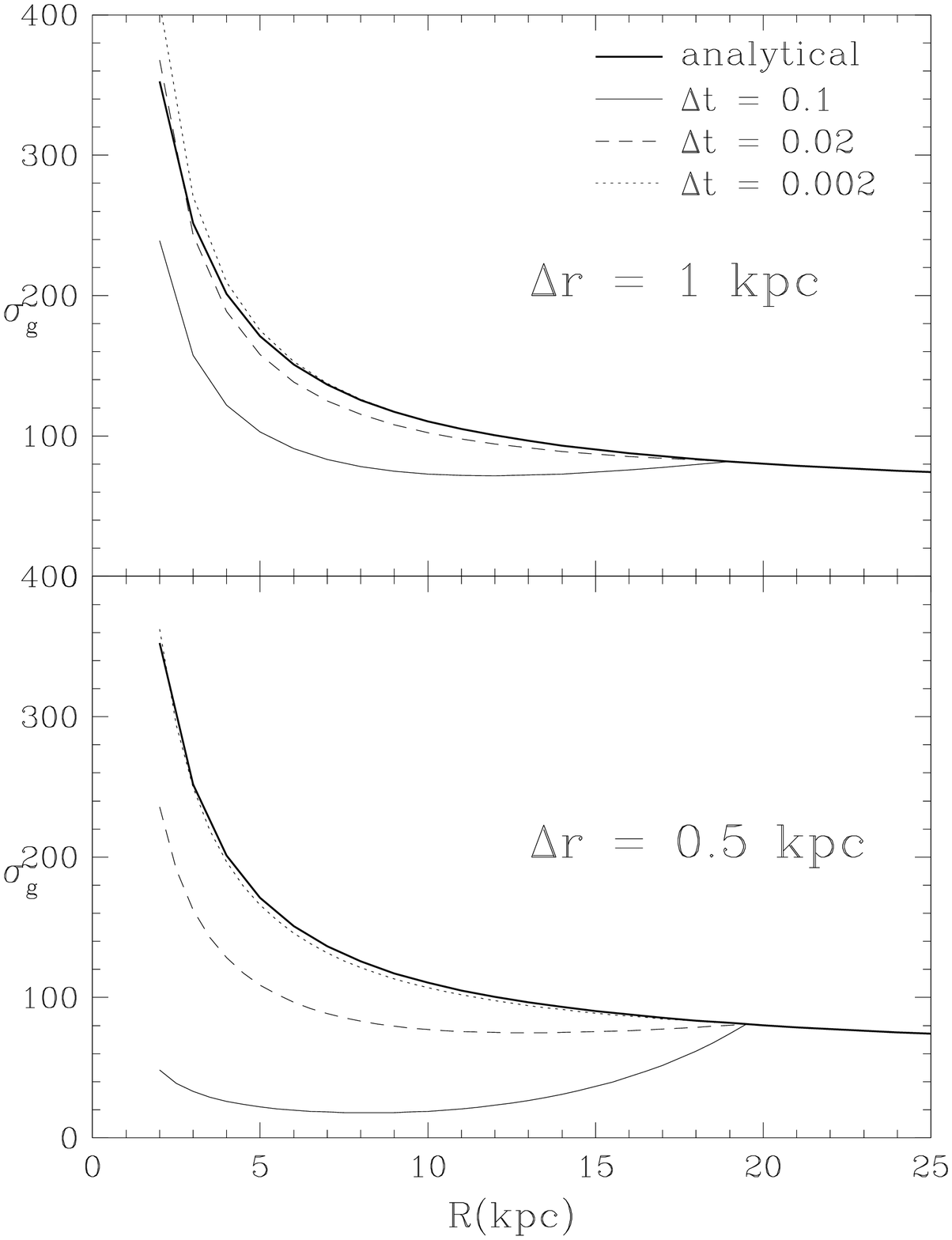,width=8.9truecm}}
\caption{Numerical models compared to the exact analytical solution for a flat
accretion profile with inflows. Different models correspond
to different typical timesteps (see legend on top right) and to different 
shell spacing (upper to lower panel).}
\label{testflatfig}
\end{figure}

\medskip
{\it Selecting the timestep}.
The first reference analytical case is that of a purely 
gaseous disc of infinite radial extent and flat profile, where the gas is:
\begin{enumerate}
\item
accreting uniformly with a time-scale $\tau$;
\item
flowing inward with a constant (in time) and uniform (in space) velocity $v$,
starting from the ``inflow onset time'' $T_{rf}$.
\end{enumerate}
Such a system is governed by the same differential 
equation~(\protect{\ref{bordereq}})
describing the adopted boundary condition at the outer disc edge in the 
chemical model (\S\protect{\ref{boundary}}). Eq.~(\protect{\ref{bordereq}}) 
is a linear, first order, partial differential equation, equivalent to the 
system:
\begin{equation}
\label{bordersystem}
\left\{ \begin{array}{l r}
\frac{dr}{dt} = v & (a) \\
 & \\
\frac{d \sigma}{dt} \,= \, A \, e^{-\frac{t}{\tau}} \,-\, \frac{v}{r} \,
\sigma & (b)
\end{array}
\right.
\end{equation}
Eq.~(\protect{\ref{bordersystem}}a) is solved as:
\[ r = v \, (t-t_0) \,+\, r_0 \]
and substitution into Eq.~(\protect{\ref{bordersystem}}b) 
yields:
\begin{equation}
\label{bordereq(b)}
\frac {d \sigma}{d t} \,+\, \frac{1}{t + p} \, \sigma \,=\, 
A e^{-\frac{t}{\tau}}~~~~~~~~~~~~~~~~~~~~~~p \equiv \frac{r_0}{v} - t_0
\end{equation}
This 
linear, first order, ordinary differential equation
is solved as:
\[ \begin{array}{l l}
\sigma(t) & = \sigma(t_0) \, e^{- \int_{t_0}^{t} \frac{1}{\xi + p} d\xi} \,+\,
          \int_{t_0}^{t} e^{- \int_{\xi}^{t} \frac{1}{\theta + p} d\theta} \,
          A \, e^{-\frac{\xi}{\tau}} d\xi\\
 & \\
          & = \sigma(t_0) \, \frac{t_0+p}{t+p} \,+\, \frac{A}{t+p} \, \tau 
	\, \times \\
 & \times \left[ (t_0+p) \, e^{-\frac{t_0}{\tau}} \,-\, 
(t+p) \, e^{-\frac{t}{\tau}}\,+\,
\tau \left( e^{-\frac{t_0}{\tau}} \,-\, e^{-\frac{t}{\tau}} \right) \right] 
\end{array} \]
Finally, replacing back:
\[ r_0 = r - v \, (t-t_0) \, ,~~~~~~~~~~~~~~~t + p = \frac{r}{v} \]
we get:
\begin{equation}
\label{bordersolv}
\begin{array}{l l}
\sigma(r,t) = & \left[ 1 - \frac{v}{r} (t-t_0) \right] \sigma(r-v(t-t_0),t_0) 
\,+\, A \, \tau \, \times \\
 & \\
\multicolumn{2}{r}{
\times \left[ \left( 1 - \frac{v}{r} (t-t_0) \right)\, e^{-\frac{t_0}{\tau}}
\,-\, e^{-\frac{t}{\tau}} \,+\, \frac{v}{r} \, \tau 
\left( e^{-\frac{t_0}{\tau}} \,-\, e^{-\frac{t}{\tau}} \right) \right]}
\end{array}
\end{equation}
Let's now set the initial conditions at $t_0$. If radial flows ``activate'' 
at a time {\mbox{$t_0=T_{rf} \geq 0$}} 
the surface density distribution for $t \leq T_{rf}$ is determined just
by the accretion profile:
\[ \sigma(r,T_{rf})=A\, \tau\, \left( 1\,-\,e^{-\frac{T_{rf}}{\tau}} \right) \]
(see Eq.~\protect{\ref{eqinfall}}) and Eq.~(\protect{\ref{bordersolv}})
becomes in fact Eq.~(\protect{\ref{borderconditionTrf}}). 
Here in our test case, Eq.~(\protect{\ref{borderconditionTrf}}) 
is the exact analytical description of the surface density profile over 
the whole disc (a part from the centre $r=0$, which is a singular point).

As a representative test, let's consider the case  {\mbox{$\tau=3$~Gyr}},
$T_{rf}=0$ and
$v=-1$~km~sec$^{-1}$.
The relevant analytical solution is plot in
Fig.~\protect{\ref{testflatfig}} for $t=t_G=15$~Gyr (thick solid line).
The numerical models used for comparison cover the radial range 2
to 20~kpc and adopt a flat accretion profile $A(r) \equiv A$. 
Their outer edge does match exactly with the analytical counterpart, since the
boundary condition at $r=20$~kpc is given by the analytical 
expression~(\protect{\ref{borderconditionTrf}}) itself.
But the predictions of numerical models at inner radii tend to deviate
from the reference density profile, and the mismatch is larger:
\begin{enumerate}
\item
the larger the typical timestep of the model;
\item
for a fixed timestep, the thinner the shells (compare upper to lower panel).
\end{enumerate}
For a typical shell width of 1~kpc, for instance, a  good match is obtained 
with model timesteps of $2 \times 10^{-2}$~Gyr, while 
model shells of
0.5~kpc an acceptable profile is obtained only with timesteps of 
$2 \times 10^{-3}$~Gyr.
Therefore, a reliable representation of radial gas flows is obtained only
with a suitably small timestep; how small, is related to the 
width of the shells, namely to the resolution of the grid spacing. 

\medskip
{\it Selecting the grid spacing.}
Since our 
models are to simulate a disc with an exponential profile,
as a second test we consider a gaseous disc with uniform and constant infall 
time-scale and inflow velocity, analogous to the previous case, but with an 
exponential profile. Namely, in the representative differential 
equation~(\protect{\ref{radfnoSF}})
the accretion profile $A(r)$ declines exponentially outward:
\[ A(r) = A(r_{\odot}) \,\, e^{-\frac{r-r_{\odot}}{r_d}} \]
Eq.~(\protect{\ref{radfnoSF}}) can then be written:
\[ \frac{\partial \sigma}{\partial t}\,+\,v\,\frac{\partial \sigma}{\partial r}
 = 
A(r_{\odot}) \, e^{\frac{r_{\odot}}{r_d}}
\, e^{-\frac{r}{r_d}} \, e^{-\frac{t}{\tau}} \,-\, 
\frac{v}{r} \,\sigma \]
%
\begin{figure}[ht]
\centerline{\psfig{file=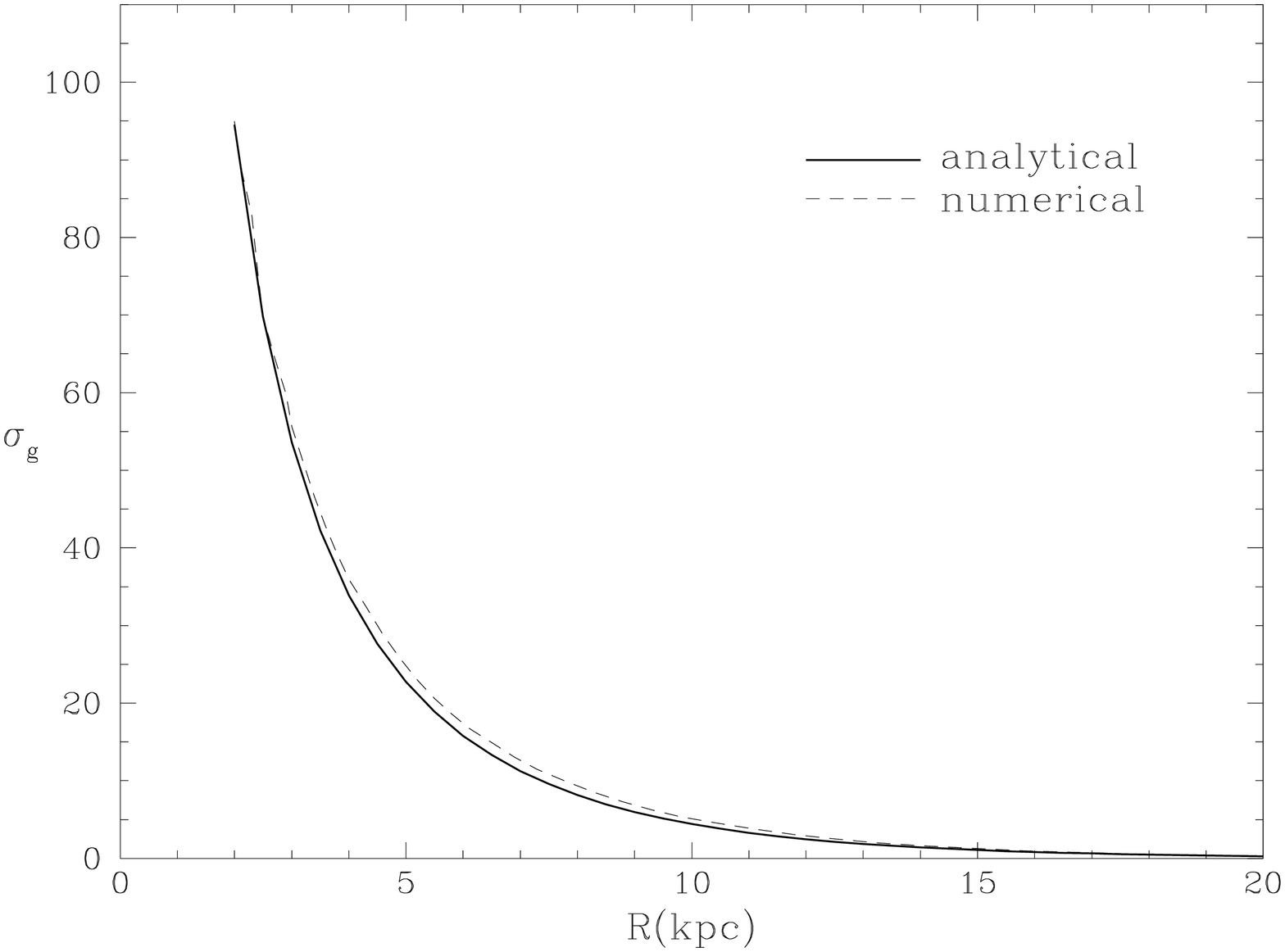,angle=-90,width=8.9truecm}}
\caption{Exact analytical solution for an exponential accretion profile 
with inflows compared to a numerical model with 40 shells 
equally spaced in logarithmic scale.}
\label{bestexpfig}
\end{figure}
%
This is another linear, first order, partial differential equation of the same
kind as (\protect{\ref{bordereq}}), and can be solved with analogous procedure
into:
\begin{equation}
\label{borderexpgen}
\begin{array}{l l}
\sigma(r,t) \, = & \left[ 1 - \frac{v}{r} (t-t_0) \right] 
\sigma(r-v(t-t_0),t_0) \,+\, \\
 & +\, \frac{ A(r_{\odot}) \, e^{-\frac{r-r_{\odot}}{r_d}}} 
{\frac{1}{\tau}+\frac{v}{r_d}} \, 
\left[ \left( 1-\frac{v}{r} (t-t_0) \right) e^{ - \frac{t_0}{\tau} +
\frac{v}{r_d}(t-t_0)} \,+ \right. \\
 & \left. -\, e^{-\frac{t}{\tau}} \,+\, 
\frac{v}{r} \frac{ e^{-\frac{t_0}{\tau}+ \frac{v}{r_d}(t-t_0)} -
e^{-\frac{t}{\tau}} }{\frac{1}{\tau}+\frac{v}{r_d}} \right]
\end{array}
\end{equation}
Notice that Eq.~(\protect{\ref{bordersolv}}) for a flat profile is recovered
from~(\protect{\ref{borderexpgen}}) for $r_d \longrightarrow \infty$. 
If radial inflows set in at time $t_0=T_{rf} \geq 0$, 
from 
\[ \sigma(r,T_{rf}) = A(r_{\odot}) \,\, e^{-\frac{r-r_{\odot}}{r_d}} \, 
\tau \, ( 1- e^{-\frac{T_{rf}}{\tau}}) \]
we get:
\begin{equation}
\label{borderexp}
\begin{array}{l l}
\sigma(r,t) & = A(r_{\odot}) \, e^{-\frac{r-r_{\odot}}{r_d}} \left\{
\tau \, \left( 1 \,-\, e^{-\frac{T_{rf}}{\tau}} \right) \, 
e^{\frac{v}{r_d} (t-T_{rf})} \,+ \right. \\
 & \left. +\, 
\frac{ e^{ \frac{v}{r_d} t - \left(\frac{1}{\tau}+\frac{v}{r_d}\right) T_{rf}} 
\,-\,  e^{-\frac{t}{\tau}} }
{\frac{1}{\tau}+\frac{v}{r_d}} \,+ \right. \\
\\
 & \left. + \frac{v}{r} \left[ \frac{ e^{ \frac{v}{r_d} t - 
\left( \frac{1}{\tau} + \frac{v}{r_d} \right) T_{rf}} - e^{-\frac{t}{\tau}}}
{\left( \frac{1}{\tau}+\frac{v}{r_d} \right)^2} \,-\, (t-T_{rf}) \times
\right. \right. \\
\multicolumn{2}{c}{  \left. \left. \times \left(
\tau \left( 1 \,-\, e^{-\frac{T_{rf}}{\tau}} \right) 
e^{\frac{v}{r_d} (t-T_{rf})} \,+\,
\frac{ e^{ \frac{v}{r_d} t - \left(\frac{1}{\tau}+\frac{v}{r_d}\right) T_{rf}}}
{\frac{1}{\tau}+\frac{v}{r_d}} 
\right) \right] \right\} } 
\end{array}
\end{equation}

We compared this second analytical case to our
numerical model, where an exponential accretion profile was adopted between 
2 and 20~kpc, and at the outer edge the boundary condition
(\protect{\ref{borderconditionTrf}}) was replaced
by~(\protect{\ref{borderexp}}).
Our tests showed that in this case the analytical profile is better 
reproduced the larger the number of shells, i.e.\ the finer the grid spacing.
A good match 
is obtained especially when model shells
are equally spaced in the logarithmic, rather than linear, scale; namely when 
the shells are chosen, in this case of an exponential accretion profile, so as
to contain roughly the same mass, rather than cover the same radial width. 
Fig.~\protect{\ref{bestexpfig}} shows in fact the analytical solution
(\ref{borderexp})
(for $r_d=4$~kpc, $\tau=3$, $T_{rf}=0$ and $v=-1$) together with a 
corresponding numerical model with 40 shells logarithmically spaced
(from 0.1~kpc wide for the inner ones to 
$\sim$1~kpc wide for the outer ones).

With such a grid spacing, our tests with the reference flat profile 
of Fig.~\protect{\ref{testflatfig}}
indicate $10^{-4}$~Gyr as a suitable timestep 
to obtain stable solutions for velocities up to the order of 
{\mbox{1~km~sec$^{-1}$}}.

In the light of all these tests, in our chemical models we adopted a
grid spacing of 35 shells from 2.5 to 20~kpc, equally spaced in 
logarithmic scale and a typical timestep 
of $10^{-4}$~Gyr (see \S\protect{\ref{numericalradf}}).
Of course, the suitable timestep depends also on the velocity field: 
flows with higher velocities require shorter integration timesteps.
Whenever we need to consider much larger speeds than {\mbox{1~km~sec$^{-1}$}},
as might be the case for the strong flows induced by the Bar, we reduce the 
timestep in proportion.



\end{document}